\documentclass[aps,superscriptaddress,showkeys]{revtex4}
\usepackage{subfigure}
\usepackage{epsfig}
\usepackage{epstopdf}
\usepackage{graphicx}
\usepackage{longtable}
\usepackage{CJK}
\usepackage{amsmath,bbold}
\usepackage{float}
\usepackage{physics}
\usepackage{comment}
\usepackage{array}

\usepackage{color}

\usepackage{hyperref}
\hypersetup{colorlinks=true, linkcolor=red, citecolor=blue,  urlcolor=magenta}

\begin{document}

\title{Coupling and Recoupling Coefficients for Wigner's U(4) Supermultiplet Symmetry}

\author{Phong Dang}
\email{pdang5@lsu.edu}
\affiliation{Department of Physics and Astronomy, Louisiana State University, Baton Rouge, LA 70803, USA}

\author{Jerry P. Draayer}
\affiliation{Department of Physics and Astronomy, Louisiana State University, Baton Rouge, LA 70803, USA}
\affiliation{Quantum CodeX, Baton Rouge, LA, USA}

\author{Feng Pan}
\affiliation{Department of Physics, Liaoning Normal University, Dalian 116029, China}
\affiliation{Department of Physics and Astronomy, Louisiana State University, Baton Rouge, LA 70803, USA}

\author{Tom\'{a}\v{s}  Dytrych}
\affiliation{Nuclear Physics Institute, Academy of Sciences of the Czech Republic, \v{R}e\v{z} 25068, Czech Republic}
\affiliation{Department of Physics and Astronomy, Louisiana State University, Baton Rouge, LA 70803, USA}

\author{Daniel Langr}
\affiliation{Department of Computer Systems, Faculty of Information Technology, Czech Technical University, Prague 16000, Czech Republic}

%\author{Kristina D. Launey}
%\affiliation{Department of Physics and Astronomy, Louisiana State University, Baton Rouge, LA 70803, USA}

\author{David Kekejian}
\affiliation{Quantum CodeX, Baton Rouge, LA, USA}

\author{Kevin S. Becker}
\affiliation{Department of Physics and Astronomy, Louisiana State University, Baton Rouge, LA 70803, USA}

%\author{Alexis Mercenne}
%\affiliation{Department of Physics and Astronomy, Louisiana State University, Baton Rouge, LA 70803, USA}

%\author{Grigor H. Sargsyan}
%\affiliation{Facility for Rare Isotope Beams, Michigan State University 640 S Shaw Ln, East Lansing, MI 48824, USA}

\author{Noah Thompson}
\affiliation{Department of Physics and Astronomy, Louisiana State University, Baton Rouge, LA 70803, USA}

%\affiliation{Department of Physics and Astronomy, Louisiana State University, Baton Rouge, LA 70803, USA}
%\author*[1]{Phong Dang}\email{pdang5@lsu.edu}

%\author[1]{Jerry P. Draayer}\email{draayer@lsu.edu}

%\author[2,1]{Feng Pan}\email{daipan@dlut.edu.cn}

%\author[1]{Kevin S. Becker}\email{kbeck13@lsu.edu}

%\affil*[1]{\orgdiv{Department of Physics and Astronomy}, \orgname{Louisiana State University}, \orgaddress{\city{Baton Rouge}, \postcode{70803-4001}, \state{LA}, \country{USA}}}

%\affil[2]{\orgdiv{Department of Physics}, \orgname{Liaoning Normal University}, \orgaddress{\city{Dalian}, \postcode{116029}, \country{China}}}

\vskip.2cm

\vskip.2cm

\begin{abstract}
\textbf{Abstract:} A novel procedure for evaluating Wigner coupling coefficients and Racah recoupling coefficients for U(4) in two group-subgroup chains is presented. The canonical $\rm U(4) \supset U(3)\supset U(2)\supset U(1)$ coupling and recoupling coefficients are applicable to any system that possesses U(4) symmetry, while the physical $\rm U(4) \supset SU_S(2) \otimes SU_T(2)$ coupling coefficients are more specific to nuclear structure studies that utilize Wigner's Supermultiplet Symmetry concept. The procedure that is proposed sidesteps the use of binomial coefficients and alternating sum series, and consequently enables fast and accurate computation of any and all U(4)-underpinned features. The inner multiplicity of a $(S,T)$ pair within a single $\rm U(4)$ irreducible representation is obtained from the dimension of the null space of the $\rm SU(2)$ raising generators; while the resolution for the outer multiplicity follows from the work of Alex et al. on $\rm U(N)$. It is anticipated that a C++ library will ultimately be available for determining generic coupling and recoupling coefficients associated with both the \textit{canonical} and the \textit{physical} group-subgroup chains of U(4).
\end{abstract}

%\abstract{...}


\maketitle

\section{INTRODUCTION}

In 1937, Wigner \cite{Wigner1937PR} followed the idea of isobaric spin -- or isospin -- proffered by Heisenberg \cite{Heisenberg1932ZP} to differentiate protons and neutrons as if they were the same particle, and proposed a unification of the spin and isospin degrees of freedom of nuclear forces into a single supermultiplet, which is described by a group-subgroup chain $\rm U(4)\supset SU_S(2) \otimes SU_T(2)$ [herein dubbed $\rm U_{ST}(4)$ for short] with $S$ and $T$ signifying the spin and isospin dependence. This theory has led to numerous influential consequences \cite{Cseh2014EPJ}, ranging from identifying ``\textit{simplicity within complexity}'' of nuclear dynamics to intriguing experimental observations. 

\vskip.2cm

The most important ramification of Wigner's Supermultiplet theory can be found in $\beta$-radioactivities \cite{Wigner1939PR} and the Gamow-Teller (GT) resonances \cite{Ikeda1962PL,Ikeda1963PL,Fujita1964PR,Fujita1965NP,Fujita2019PRC}. Since the GT transition operator is a generator of the U(4) group, it does not couple states belonging to different U(4) irreducible representations (hereafter referred to as irreps), $\beta$-decay only occurs within a single U(4) supermultiplet. By investigating the GT resonances in a wide range of nuclei, Lutostanky and collaborators showed that the U(4) Supermultiplet Symmetry is restored for neutron-rich heavy and superheavy nuclei, based on a decrease in spin-orbit splitting \cite{Gaponov2010PoAN,Lutostansky2016EPJ}, which is of great ongoing interest for modelling and understanding stellar nucleosynthesis processes that underpin the creation of the Universe.

\vskip.2cm

The existence of Wigner's $\rm U_{ST}(4)$ symmetry also seems to emerge from the residual strong force between the nucleons themselves. In 1996, Kaplan and Savage proved that the $\rm U_{ST}(4)$ symmetry is an accidental (or emergent) symmetry of the leading-order operators belonging to a chiral Lagrangian at the large-$N_c$ limit of QCD \cite{Kaplan1996PLB}, followed by a deeper analysis of the $NN$ potential that imitates two-body nuclear forces \cite{Kaplan1997PRC}. The same spirit was also carried out in a framework of one boson exchange potential \cite{Cordon2008PRC} and another of large scattering lengths \cite{Mehen1999PRL,Beane2013PRC}. More recently, with the development of lattice QCD calculations, Lu et al. \cite{Lu2019PLB} showed from a $\rm U_{ST}(4)$-symmetric Hamiltonian with up to three-body interactions that the most essential elements of nuclear forces are those that satisfy $\rm U_{ST}(4)$ invariance. In addition, the advent of quantum entanglement and information theory has also shed new lights on the origin of the $\rm U_{ST}(4)$ symmetry within the context of entanglement suppression of the strong interactions \cite{Beane2019PRL,Liu2023PRC,Miller2023PRC}.

\vskip.2cm

It now seems timely to suggest that the Supermultiplet theory proffered by Wigner nearly a century ago has proven itself to be of extreme importance in understanding the dynamics of nuclear matter at low energies. Although some studies have been carried out from a shell-model perspective \cite{Kota2024PScr}, there has not been a thorough utilization of the $\rm U_{ST}(4)$ symmetry that is on par with what has been done for the $\rm SU(3)$ symmetry proposed by Elliott in 1958 \cite{Elliott1958PRSA,Elliott1958PRSA-II}; see applications of the SU(3) symmetry in, for example, connection to the Liquid Drop Model \cite{Draayer1989PRL}, the Symplectic Shell Model \cite{Rosensteel1977AnnP,Rosensteel1980AnnP}, the Symmetry-adapted No-core Shell Model (SA-NCSM) \cite{Dytrych2007PRL,Dytrych2020PRL}, the Multiconfigurational Dynamical Symmetry (MUSY) framework \cite{Cseh2021PRC}, the Pseudo- and Proxy-SU(3) Schemes \cite{Raju1973NPA,Bonatsos2017PRC,Cseh2020PRC}, etc. The main culprit behind the lack of the former is the branching multiplicity issues that occur in the reduction from the group $\rm U(4)$ to the two subgroups $\rm SU_S(2)$ and $\rm SU_T(2)$, and in the coupling of different $\rm U(4)$ irreps. Thanks to the contemporary work on Wigner's Supermultiplet theory \cite{Pan2023EPJP,Pan2023NPA,Pan2024CPC} and more generally on $\rm U(N)$ \cite{Alex2011JMP}, it is now possible to construct a next-generation robust symmetry-adapted shell model theory, one where spatial and spin-isospin degrees of freedom are managed equitably with one another. 

\vskip.2cm 

The construction of such a comprehensive theory requires the evaluation of coupling and recoupling coefficients for both $\rm SU(3)$ and $\rm U_{ST}(4)$. For the former, a Fortran package was published more than 50 years ago \cite{Akiyama1973CPC,Draayer1973JMP}, which was then converted into C++ \cite{Dytrych2021CPC} using the same algorithm but with boosted precision. Recently, a new procedure for evaluating coupling and recoupling coefficients for $\rm U(3)$ has been proposed by Dang, et al. \cite{Dang2024}, wherein the resolution of the outer and inner multiplicities was based on the nullspace concept of the $\rm U(3)$ generators \cite{Alex2011JMP,Pan2016NPA}. Hecht and Pang first attempted to tackle the $\rm U_{ST}(4)$ side in 1969 \cite{Hecht1969JMP}, and then Draayer independently worked on the transformation between the canonical and physical bases of $\rm U(4)$ \cite{Draayer1970JMP}. There were also other articles by Partensky and Maguin \cite{Partensky1978JMP}, and by Rowe and Repka \cite{Rowe1997FoP}, which give explicit expression for some specific Wigner $\rm U_{ST}(4)$ coupling and recoupling coefficients. Nevertheless, none of these provided a simple, yet rigorous procedure for the computation of arbitrary coupling and recoupling coefficients of $\rm U_{ST}(4)$. This paper, therefore, serves to bring forward a simple algorithm for the evaluation of any and all coefficients that are required for modern shell-model calculations, upon which an early version of a C++ library has been built that demonstrates promising performance. 

\vskip.2cm 

%In what follows, Section \ref{secII} is dedicated to a brief review of the unitary group $\rm U(4)$ and its irreps, which is followed by Section \ref{secIII} to establish the labelling conventions for the canonical group-subgroup chain of U(4), in particular, the labelling of basis states and how they transform under the actions of the generators of U(4). Section \ref{secIV} lays out a procedure for calculating the $\rm U(4) \supset U(3) \supset U(2) \supset U(1)$ canonical coupling coefficients based upon the more universal methodology for $\rm U(N)$ given by Alex et al. \cite{Alex2011JMP}, \red{which are shown in Section \ref{secV} to be useful for computing the recoupling coefficients that are independent of the group-subgroup structure. Thereafter comes Section \ref{secVI}, which gives an outline on how Wigner coupling coefficients for $\rm U_{ST}(4)$ are calculated \cite{Pan2023NPA} with a projection procedure onto the canonical basis as introduced by Pan et al. \cite{Pan2023EPJP}. Lastly, some concluding remarks are proffered in Section \ref{secVII}. }

In what follows, Section \ref{secII} is dedicated to a brief review of the unitary group $\rm U(4)$ and its irreps. The latter is followed by Section \ref{secIII} which serves to establish the labelling conventions for the canonical group-subgroup chain of U(4), and how they transform under the action of the generators of U(4). Section \ref{secIV} lays out a procedure for calculating the $\rm U(4) \supset U(3) \supset U(2) \supset U(1)$ canonical coupling coefficients based upon the methodologies for $\rm U(N)$ proffered by Alex et al. \cite{Alex2011JMP}, which in turn are shown in Section \ref{secV} to be useful for computing the recoupling coefficients that are independent of the group-subgroup structure. Section \ref{secVI} gives a description on how Wigner coupling coefficients for $\rm U_{ST}(4)$ are calculated \cite{Pan2023NPA} with a projection procedure onto the canonical basis as introduced by Pan et al. \cite{Pan2023EPJP}. And lastly, Section \ref{secVII} is reserved for some concluding comments.

\section{The Unitary Group U(4) and its Irreducible Representations}
\label{secII}

The unitary group U(4) is a rank-3 group with sixteen generators: $E_{ij}$ (where $i,j=1,2,3,4$) which obey the following commutation relation and conjugation property:
\begin{align}
    [E_{ij},E_{mn}] = \delta_{jm}E_{in} - \delta_{in}E_{mj} \text{ and } E_{ij}^\dagger &= E_{ji},
    \label{eq:properties}
\end{align}
where $\delta_{ij}$ is the standard Kronecker delta symbol. Various subgroups of U(4), and hence group-subgroup chains, can be realized by taking linear combinations of these operators in different ways. The most obvious group-subgroup structure is:
\begin{align}
    \rm U(4) = U(1) \otimes SU(4),
\end{align}
where the subgroup SU(4) is a special unitary group composed of fifteen generators that remain after removing the first order Casimir invariant of U(4),  $C_1[{\rm U(4)}] = \sum_{i=1}^4 E_{ii}$, which is essentially the trace of a four-dimensional matrix representation of the sixteen generators of the U(4) group. 
\vskip.2cm
\begin{figure}[h!]
    \centering
    \includegraphics[scale=0.4]{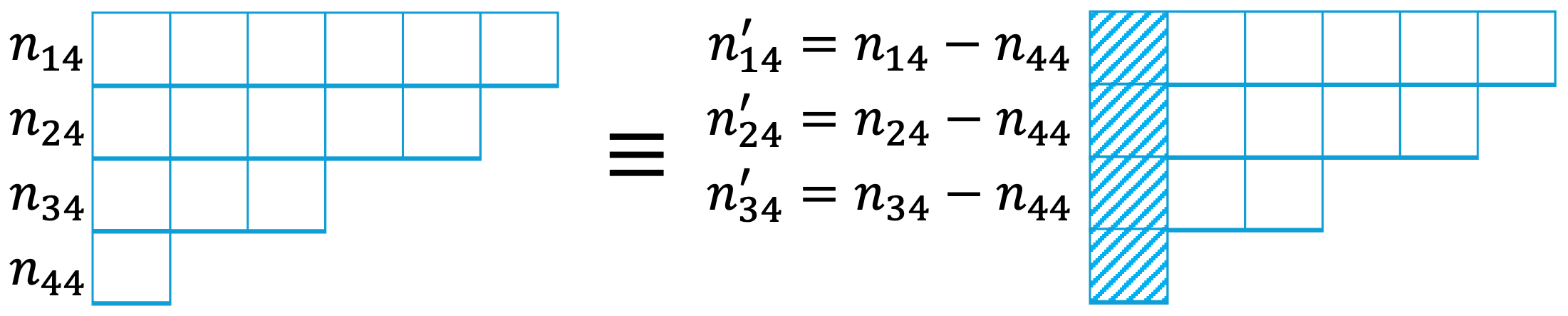}
    \caption{Young tableau representation of U(4) [on the left] and SU(4) [on the right]  irreps.}
    \label{fig:Young-tab}
\end{figure}

An irrep of U(4) is labeled by four non-negative integers, $[n_{14},n_{24},n_{34},n_{44}]$, which are by definition in decreasing order, i.e., $n_{14}\ge n_{24} \ge n_{34} \ge n_{44} \ge 0$, and are often understood as the maximum numbers of particles in the four spin-isopsin single-particle states ($\ket{\uparrow\uparrow}$, $\ket{\uparrow\downarrow}$, $\ket{\downarrow\uparrow}$, $\ket{\downarrow\downarrow}$).  Since two U(4) irreps $[n_{14},n_{24},n_{34},n_{44}]$ and $[n_{14}+\nu,n_{24}+\nu,n_{34}+\nu,n_{44}+\nu]$ are equivalent and have the same dimension (whose formula will be given in the next section), a SU(4) irrep can be labeled by one less quantum number $[n_{14}'=n_{14}-n_{44},n_{24}'=n_{24}-n_{44},n_{34}'=n_{34}-n_{44}]$, where $[n_{14},n_{24},n_{34},n_{44}]$ is a associated U(4) irrep. (Note that in the work of Alex et al. \cite{Alex2011JMP}, this action is termed as ``normalizing" an irrep.) From this, it can observed that a SU(4) irrep corresponds to multiple U(4) irreps, which implies that the contraction from U(4) to SU(4) results in a loss of ``information" about the total number of particles that are being distributed among the four single-particle states. However, as the relative distribution between them is still known via the SU(4) labels, the full accounting of the total number of particles can be restored if the value of the first-order Casimir invariant of U(4) is being kept track of. One can also use the so-called Young tableau as a graphic representation for the irreps of U(4) and SU(4), which subsumes four rows for the former and three rows for latter, wherein the number of boxes on each row is equal to the corresponding partition quantum number of the irrep, see FIG. \ref{fig:Young-tab}. Briefly stated, for any U(4) tableau representation, the corresponding SU(4) counterpart can be obtained by simply removing the full columns appearing on the left-hand side of the U(4) tableau.

\section{The Canonical Group-Subgroup Chain U(4)$\supset$U(3)$\supset$U(2)$\supset$U(1)}
\label{secIII}

\begin{figure}[h!]
    \centering
\includegraphics[scale=0.6]{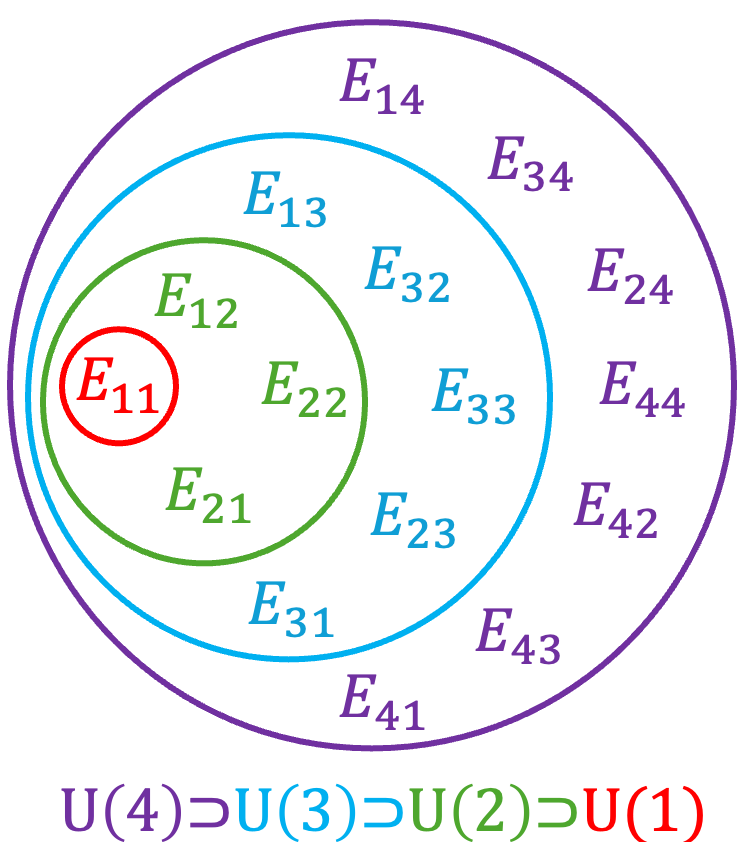}
    \caption{A schematic depiction of the nine generators of the canonical $\rm U(4) \supset U(3) \supset U(2) \supset U(1)$ group-subgroup chain.}
    \label{fig:canonical_chain}
\end{figure}

Among the sixteen generators, $E_{ij}$ ($i,j=1,2,3,4$), of the U(4) group, there are nine that span the U(3) algebra; specifically, this includes the subset $E_{ij}$ ($i,j=1,2,3$), out of which four operators, $E_{ij}$ ($i,j=1,2$), generate the infinitesimal transformations of the group U(2), and finally, $E_{11}$ is the only generator of the U(1) group, see FIG. \ref{fig:canonical_chain} for a graphic illustration of this group-subgroup structure. Hence, the reduction $\rm U(4) \supset U(3) \supset U(2) \supset U(1)$ is a natural group-subgroup chain of the U(4) group, which is commonly referred to as the canonical chain that has been studied extensively in the literature, for example see \cite{Louck1970JMP,Biedenharn1972JMP1,Biedenharn1972JMP2,Louck1973JMP,Kuhn2008CPC,Alex2011JMP}. 
 
\begin{figure}
    \centering
    \includegraphics[scale=0.7]{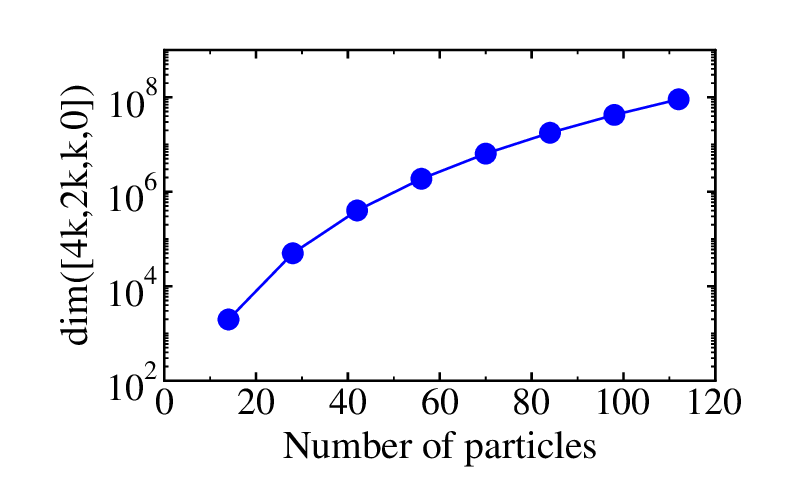}
    \caption{Dimension of some U(4) irreps of the type $[4k,2k,k,0]$ (with even $k$), where $A=\sum_i n_{i4} = 7k$ is the total number of particles that are distributed into the irreps.}
    \label{fig:dim}
\end{figure}

The basis states of this canonical scheme can be written in an elegant pattern known as a \textit{Gel'fand-Tsetlin pattern} \cite{Gelfand1950} -- hereafter called a Gelfand state -- which is composed of ten quantum numbers organized in a four-row triangular shape as follows,
\begin{equation*}
\ket{G} = \Bigg|
\begin{array}{c}
n_{14}, n_{24}, n_{34}, n_{44} \\
n_{13}, n_{23}, n_{33} \\
n_{12}, n_{22} \\
n_{11}
\end{array}
\Bigg\rangle
\end{equation*}
Within this simple graphical representation, the so-called \textit{betweenness} conditions, $ n_{ij} \ge n_{i,j-1} \ge n_{i+1,j}$,  must hold. For a given U(4) irrep, these betweenness conditions uniquely determine the complete set of Gelfand states that are associated with the irrep. This set of all Gelfand states is known as the carrier space of that irrep; and the total number of the states therein is by definition the dimension of the irrep, see FIG. \ref{fig:dim}, which can be computed from the quantum numbers of the irrep via the following formula:
\begin{small}
\begin{multline}
    \dim([n_{14},n_{24},n_{34},n_{44}]) = \frac{1}{12}(1+n_{14}-n_{24})(2+n_{14}-n_{34})(3+n_{14}-n_{44})(1+n_{24}-n_{34})(2+n_{24}-n_{44})(1+ n_{34}-n_{44}).
    \label{eq:dim1}
\end{multline}
\end{small}

\vskip.2cm

As pointed out earlier in Section \ref{secII}, a SU(4) irrep is simply achieved by subtracting the last quantum number $n_{44}$ from the others (i.e., $n_{14},n_{24},n_{34}$) of a U(4) irrep, which implies an exact correspondence under which the matrix representations associated with a SU(4) irrep and its corresponding U(4) counterpart are the same: 
\begin{align}
\Bigg| 
\begin{array}{c}
n_{14},n_{24},n_{34},n_{44}\cr
n_{13},n_{23},n_{33}\cr
n_{12},n_{22}\cr
n_{11}
\end{array}
\Bigg\rangle = \Bigg| \begin{array}{c}
n_{14}-n_{44},n_{24}-n_{44},n_{34}-n_{44},0\cr
n_{13}-n_{44},n_{23}-n_{44},n_{33}-n_{44}\cr
n_{12}-n_{44},n_{22}-n_{44}\cr
n_{11}-n_{44}
\end{array}
\Bigg\rangle.
\end{align}

\vskip.2cm

Although the generators of the U(4) group only act within the carrier space of a single irrep, their action on different Gelfand states within that carrier space generally varies. However, there are four special cases where the generators do not change the Gelfand states that are acted upon, namely,
\begin{align}
    E_{11}\ket{G} &= n_{11} \ket{G}, \nonumber\\
    E_{22}\ket{G} &= (n_{12} + n_{22} - n_{11}) \ket{G},     \label{eq:remain}\\
    E_{33}\ket{G} &= (n_{13}+n_{23}+n_{33}-n_{12}-n_{22}) \ket{G}, \nonumber\\
    E_{44}\ket{G} &= (n_{14}+n_{24}+n_{34}+n_{44}-n_{13}-n_{23}-n_{33}) \ket{G}. \nonumber
\end{align}
These operators are known as the diagonal generators, and their function is simply to count the numbers of quanta in the basis state that are distributed in the four single-particle states. It is also important to note that all of the Gelfand states are eigenstates of the diagonal generators, and the corresponding eigenvalues constitute a so-called pattern weight vector (or p-weight for short) \cite{Alex2011JMP}: $[w_4,w_3,w_2,w_1]:=[n_{14}+n_{24}+n_{34}+n_{44}-n_{13}-n_{23}-n_{33},n_{13}+n_{23}+n_{33}-n_{12}-n_{22},n_{12}+n_{22}-n_{11},n_{11}]$. It is important to note that a Gelfand state cannot be uniquely identified by a p-weight; indeed, the same p-weight can belong to multiple Gelfand states. Moreover, as the U(4) quantum numbers $n_{14},n_{24},n_{34},n_{44}$ are already specified, an alternative weight with only three numbers, $[n_{13}+n_{23}+n_{33},n_{12}+n_{22},n_{11}]$ can be used, which is known as the z-weight vector of a Gelfand state, (for more details about the definition of the z-weight and its one-to-one correspondence to the p-weight, see \cite{Alex2011JMP}). 

\vskip.2cm
\begin{figure}[h!]
    \centering
    \includegraphics[scale=0.7]{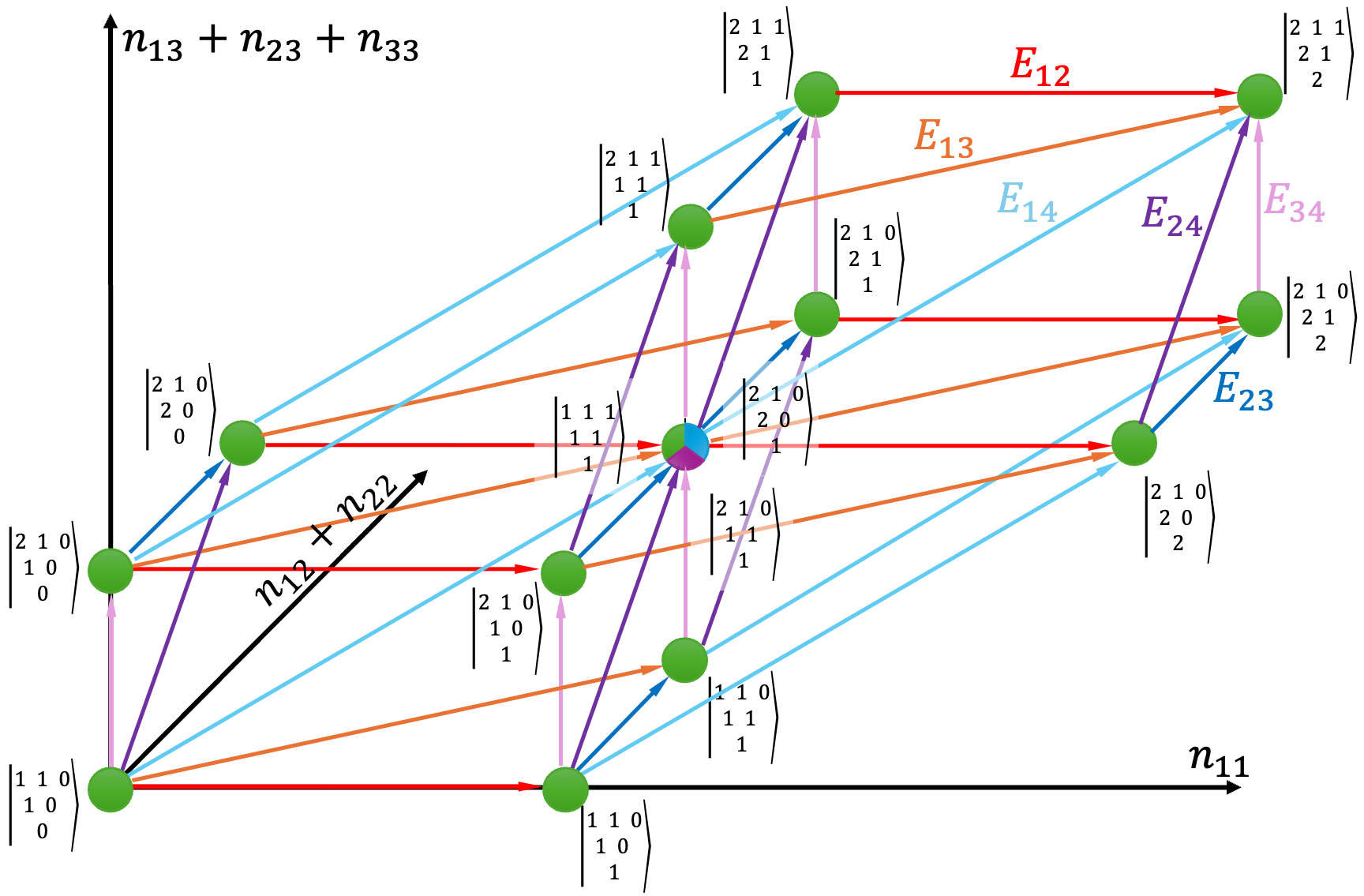}
    \caption{The z-weight diagram for the U(4) irrep $[2,1,1,0]$, with a demonstration of the action -- using colored arrows -- on all of the Gelfand states of the U(4) raising generators that push upward towards the so-called highest weight state (sometimes referred to as the extremal state) at the top right corner of the plot. The conjugate of the highest weight state is the lowest weight state located at the origin at the bottom left corner. Since the irrep is specified, the top row of the Gelfand patterns are omitted.}
    \label{fig:generators}
\end{figure}

\vskip.2cm

In contrast with the above diagonal generators, the off-diagonal generators do connect different Gelfand states within the carrier space, which can be categorized into two groups of \textit{raising} and \textit{lowering} operators. In what follows, for convenience, a shorthand notation is introduced, $\ket{G \pm I_{ij}}$, which simply means that the entry $n_{ij}$ of the Gelfand pattern $\ket{G}$ that is being affected is either added by 1 (for plus sign) or subtracted by 1 (for minus sign). (It is important not to mistake this notation for an addition or a subtraction between the Gelfand states.) With this schematic notation in play the action of the six raising generators is as follows:

\begin{align}
    E_{12}\ket{G} &= e_{12}(G) \ket{G+I_{11}}, \nonumber\\
    E_{23}\ket{G} &= e_{23,12}(G) \ket{G+I_{12}} + e_{23,22}(G) \ket{G+I_{22}}, \nonumber \\
    E_{13}\ket{G} &= e_{13,12}(G) \ket{G+I_{12}+I_{11}}+e_{13,22}(G) \ket{G + I_{22} + I_{11}}, \nonumber\\
    E_{34}\ket{G} &= e_{34,13}(G) \ket{G+I_{13}} + e_{34,23}(G) \ket{G+I_{23}} + e_{34,33}(G) \ket{G+I_{33}}, \nonumber \\
    E_{24}\ket{G} &= e_{24,13,12} \ket{G+I_{13}+I_{12}} + e_{24,13,22} \ket{G+I_{13}+I_{22}} + e_{24,23,12} \ket{G+I_{23}+I_{12}} \label{eq:raising}\\
        &+ e_{24,23,22} \ket{G+I_{23}+I_{22}} + e_{24,33,12} \ket{G+I_{33}+I_{12}} + e_{24,33,22} \ket{G+I_{33}+I_{22}}, \nonumber \\
    E_{14}\ket{G} &= e_{14,13,12} \ket{G+I_{13}+I_{12}+I_{11}} + e_{14,13,22} \ket{G+I_{13}+I_{22}+I_{11}} + e_{14,23,12} \ket{G+I_{23}+I_{12}+I_{11}} \nonumber\\
        &+ e_{14,23,22} \ket{G+I_{23}+I_{22}+I_{11}} + e_{14,33,12} \ket{G+I_{33}+I_{12}+I_{11}} + e_{14,33,22} \ket{G+I_{33}+I_{22}+I_{11}}. \nonumber
\end{align}
In the above expressions the coefficients in front of the new states are given in Appendix \ref{appendA}. For the lowering generators, one can easily deduce and visualize their action based upon the conjugation property that $E_{ij}^\dagger = E_{ji}$:
\begin{align}
    E_{21}\ket{G} &= e_{21}(G) \ket{G-I_{11}}, \nonumber\\
    E_{32}\ket{G} &= e_{32,12}(G) \ket{G-I_{12}} + e_{32,22}(G) \ket{G-I_{22}}, \nonumber \\
    E_{31}\ket{G} &= e_{31,12}(G) \ket{G-I_{12}-I_{11}}+e_{31,22}(G) \ket{G-I_{22}-I_{11}}, \nonumber\\
    E_{43}\ket{G} &= e_{43,13}(G) \ket{G-I_{13}} + e_{43,23}(G) \ket{G-I_{23}} + e_{43,33}(G) \ket{G-I_{33}}, \nonumber \\
    E_{42}\ket{G} &= e_{42,13,12} \ket{G-I_{13}-I_{12}} + e_{42,13,22} \ket{G-I_{13}-I_{22}} + e_{42,23,12} \ket{G-I_{23}-I_{12}} \label{eq:lowering}\\
        &+ e_{42,23,22} \ket{G-I_{23}-I_{22}} + e_{42,33,12} \ket{G-I_{33}-I_{12}} + e_{42,33,22} \ket{G-I_{33}-I_{22}}, \nonumber \\
    E_{41}\ket{G} &= e_{41,13,12} \ket{G-I_{13}-I_{12}-I_{11}} + e_{41,13,22} \ket{G-I_{13}-I_{22}-I_{11}} + e_{41,23,12} \ket{G-I_{23}-I_{12}-I_{11}} \nonumber\\
        &+ e_{41,23,22} \ket{G-I_{23}-I_{22}-I_{11}} + e_{41,33,12} \ket{G-I_{33}-I_{12}-I_{11}} + e_{41,33,22} \ket{G-I_{33}-I_{22}-I_{11}}, \nonumber
\end{align}
where the coefficients can be found in Appendix \ref{appendB} of this article. Figure \ref{fig:generators} is a z-weight diagram that serves as a demonstration for the action of the raising generators on the Gelfand basis states that constitute the carrier space of the U(4) irrep $[2,1,1,0]$. Note that at the center of the diagram are three states of the same weight, which are shown by three colors. Although the action of the lowering generators are not explicitly displayed here, they can be visualized via arrows that point in the opposite direction to those shown in Figure \ref{fig:generators} for the raising generators.

\section{U(4)$\supset$U(3)$\supset$U(2)$\supset$U(1) Coupling Coefficients}
\label{secIV}

The need for coupling group irreps arises within the broader context of seeking to unveil the true nature of composite systems that seem to be reflective of special symmetries that are found to be resident, and often dominant, within the constituent parts of the system under consideration. The simplest case to consider is the coupling of two U(4) irreps, e.g., binary clustering of atomic nuclei \cite{Cseh1992PLB,Cseh1994AnnP,Dang2023PRC}, to each of which there is a unique carrier space associated. In most scenarios, the coupling leads to a variety of irreps, which also have their own carrier spaces and each may occur more than once, hence one needs to acquire the knowledge about the transformation from the uncoupled spaces to the coupled ones and vice versa. This transformation is given via the so-called Clebsch-Gordan coefficients (abbreviated as CGC's), and for the SU(2) algebra of total angular momentum, the CGC's are $\braket{j_1m_1;j_2m_2}{JM}$, or the equivalent 3-$j$ symbols introduced by Wigner. In an analogous manner, when one works with the U(4) group, the CGC's provide the expansion coefficients of a Gelfand state in the coupled carrier space in terms of those in the uncoupled carrier spaces as follows,
\begin{align}
    \ket{G''}_\eta = \sum_{G,G'} \braket{G;G'}{G''}_\eta \ket{G}\ket{G'} = \sum_{G,G'} C_{G,G'}^{G''_\eta} \ket{G}\ket{G'},
    \label{eq:cano-expansion}
\end{align}
where $C_{G,G'}^{G''_\eta}$ are the canonical $\rm U(4) \supset U(3)\supset U(2) \supset U(1)$ CGC's, which for the matter of convention we call Wigner 3-U(4) coefficients. Here the integer $\eta$ is simply a counter for multiple occurrences of $\ket{G''}$, which is discussed in more details below. Note that the formulation in this section is derived explicitly for U(4) from a more general procedure for U(N) brought forward by Alex et al. \cite{Alex2011JMP}, and as well resembles a recent work on U(3) \cite{Dang2024}.

\subsection{Direct Product of Two U(4) Irreps}
\label{secIVA}

The direct product of two U(4) irreps, $[n_{14},n_{24},n_{34},n_{44}]$ and $[n_{14}',n_{24}',n_{34}',n_{44}']$, is in general reducible into a direct sum of U(4) irreps, each of which may occur multiple times:
\begin{align}
    [n_{14},n_{24},n_{34},n_{44}] \otimes [n_{14}',n_{24}',n_{34}',n_{44}'] = \oplus_{\eta=1}^{\eta_{\max}} [n_{14}'',n_{24}'',n_{34}'',n_{44}''], 
\end{align}
where the outer multiplicity $\eta_{\max}$ keeps track of the total number of times that the coupled irrep $[n_{14}'',n_{24}'',n_{34}'',n_{44}'']$ shows up in the decomposition. (Even though it is not a focus of this article, the authors would like to note that it may be interesting to investigate the physical significance of the outer multiplicity in future works.) The procedure to obtain the coupled irreps and the outer multiplicities thereof is well-known and often based on coupling the Young tableaux representing the irreps. For computer implementation, it is convenient to utilize the general Littlewood-Richardson rule for U(N), which is described in details in section VIII of \cite{Alex2011JMP} with an explicit example for U(3). 

\vskip.2cm

As noted in Section \ref{secIII}, the actions of all generators $E_{ij}$ are only defined within the unique carrier space of a single irrep. Therefore, as two irreps are coupled together to obtain a third one, it is important to know the relationship between the generators acting within the carrier spaces of the former and those acting within the carrier space associated with the latter as follows
\begin{align}
    E_{ij}'' = E_{ij}\otimes \mathbf{1}' + \mathbf{1}\otimes E_{ij}',
\end{align}
where $\mathbf{1}$ is the identity operator. One can verify easily that the generators $E_{ij}''$, if constructed in this way, follow all of the commutation and conjugation properties given in Eq. (\ref{eq:properties}).

\subsection{Selection Rule for U(4)$\supset$U(3)$\supset$U(2)$\supset$U(1) CGC's and Orthogonality Relations}
\label{secIVB}

By applying the diagonal generators $E_{ii}'' = E_{ii}\otimes\mathbf{1}' + \mathbf{1}\otimes E_{ii}'$ (with $i=1,2,3,4$) on  both sides of Eq. (\ref{eq:cano-expansion}), one arrives at
\begin{align}
    w_i'' \ket{G''}_\eta = \sum_{G,G'} C_{G,G'}^{G''_\eta} (w_i + w_i')\ket{G}\ket{G'},
\end{align}
which implies that $w_i'' = w_i + w_i'$. (Recall that the $[w_4,w_3,w_2,w_1]$ quartet is the p-weight vector of a Gelfand state.) Consequently, these conditions pose a constraint on which Gelfand states $\ket{G}$ and $\ket{G'}$ appear in the expansion of $\ket{G''}$, see Eq. (\ref{eq:cano-expansion}). In other words, there is a selection rule for the CGC's, $C_{G,G'}^{G''_\eta}$, of U(4) as follows:
\begin{align}
    C_{G,G'}^{G''_\eta} = 0 \text{ unless } w_i'' = w_i+w_i' \text{ with } i=1,2,3,4.
\end{align}
Moreover, since the condition $w_4''=w_4+w_4'$ is already guaranteed in the coupling of irreps, it suffices to check the other three conditions when expanding $\ket{G''}$ in the uncoupled basis $\ket{G}\ket{G'}$, which can be simplified to three equalities
\begin{align}
    n_{11}'' &= n_{11} + n_{11}', \nonumber\\
    n_{12}'' + n_{22}'' &= n_{12} + n_{22} + n_{12}' + n_{22}', \\
    n_{13}'' + n_{23}'' + n_{33}'' &= n_{13}+n_{23}+n_{33} + n_{13}'+n_{23}'+n_{33}', \nonumber
\end{align}
which is also in consistency with the rank 3 of the U(4) group. 

Also, as the CGC's are real, they should satisfy the following orthogonality relations
\begin{equation}
    \sum_{G,G'} C_{G,G'}^{G''_{1,\eta_1}} \times C_{G,G'}^{G''_{2,\eta_2}} = \delta_{\eta_1,\eta_2} \delta_{G''_1,G''_2} \text{ and } \sum_{\eta,G''} C_{G_1,G'_1}^{G'',\eta} \times C_{G_2,G'_2}^{G'',\eta} = \delta_{G_1,G_2} \delta_{G'_1,G'_2}.
\end{equation}

\subsection{U(4)$\supset$U(3)$\supset$U(2)$\supset$U(1) CGC's for the Highest Weight States}
\label{secIVC}

The upper limit of the betweenness condition (Section \ref{secIII}) defines the highest weight state -- or the stretched state -- for a given U(4) irrep $[n_{14},n_{24},n_{34},n_{44}]$, i.e.,
\begin{align}
    \ket{HW} := \Bigg| \begin{array}{c}
n_{14},n_{24},n_{34},n_{44}\cr
n_{14},n_{24},n_{34}\cr
n_{14},n_{24}\cr
n_{14}
\end{array}
\Bigg\rangle,
\end{align}
which is annihilated via the action of all six of the raising generators $E_{12}$, $E_{23}$, $E_{13}$, $E_{34}$, $E_{24}$, and $E_{14}$, see Eq. (\ref{eq:raising}) and FIG. \ref{fig:generators}. Hence, the highest weight state of a given irrep belongs to the intersection of the null spaces of those six generators. Nonetheless, only the three generators $E_{12}$, $E_{23}$ and $E_{34}$ are completely independent due to the commutation relations, Eq. (\ref{eq:properties}), thus any state that simultaneously belongs to the null space of those three is guaranteed to belong to the null space of the rest. 

\vskip.2cm

Let us apply $E_{i,i+1}''=E_{i,i+1}\otimes \mathbf{1}' + \mathbf{1}\otimes E_{i,i+1}'$ (with $i=1,2,3$) on both sides of Eq. (\ref{eq:cano-expansion}) wherein $\ket{G''}_\eta = \ket{HW''}_\eta$, one can obtain the following set of homogeneous linear equations:
\begin{small}
\begin{multline}
    \sum_{G,G'} C_{G,G'}^{HW''_\eta} \left(e_{12}(G) \ket{G+I_{11}}\ket{G'} + e_{12}(G') \ket{G}\ket{G'+I_{11}} \right) = 0,  \\
    \sum_{G,G'} C_{G,G'}^{HW''_\eta} \biggl( e_{23,12}(G) \ket{G+I_{12}}\ket{G'} 
            + e_{23,22}(G) \ket{G+I_{22}}\ket{G'} + e_{23,12}(G') \ket{G}\ket{G'+I_{12}} + e_{23,22}(G') \ket{G}\ket{G'+I_{22}} \biggl) =0,  \\
    \sum_{G,G'} C_{G,G'}^{HW''_\eta} \biggl( e_{34,13}(G) \ket{G+I_{13}}\ket{G'} 
            + e_{34,23}(G) \ket{G+I_{23}}\ket{G'} + e_{34,33}(G) \ket{G+I_{33}}\ket{G'} + e_{34,13}(G') \ket{G}\ket{G'+I_{13}}  \\
            + e_{34,23}(G') \ket{G}\ket{G'+I_{23}} + e_{34,33}(G) \ket{G}\ket{G'+I_{33}} \biggl) =0. 
\end{multline}
\end{small}

Since we know all states that are results of the action of the raising generators, Eq. (\ref{eq:raising}), these linear equations can be written in a matrix form
\begin{align}
    \mathbf{P}(HW'')\mathbf{C}^\eta = \mathbf{0},
\end{align}
where the CGC's $C_{G,G'}^{HW''_\eta}$ are entries of the column vectors $\mathbf{C}^\eta$ (for all $\eta = 1,2,...,\eta_{\max}$), and elements of the matrix $\mathbf{P}(HW'')$ can be computed from the coefficients in Eq. (\ref{eq:raising}) and Appendix \ref{appendA}. Then it is straightforward to obtain the CGC's simultaneously for all values of $\eta$ by solving the null space of the matrix $\mathbf{P}(HW'')$, which can be done easily using various numerical methods that are now available; and the dimension of the null space must be equal to the outer multiplicity $\eta_{\max}$. In case the solution of the null space does not exist, the coupling of the two irreps is simply not permitted. When there is only one solution, the set of coupling coefficients to the highest weight state is uniquely defined. However, it often occurs that the outer multiplicity of the coupled irrep is larger than one, in which case the CGC's no longer uniquely exist and are contingent upon the numerical method which is employed to solve the null space \cite{Alex2011JMP}. Despite this ambiguity, the solutions based on different numerical methods are equivalent and can be realized from one another by a unitary or orthogonal transformation. It is also important to note that the numerical solution of the null space may not contain orthonormal column vectors; should that be the case, one needs to carry out a Gram-Schmidt procedure to attain a set of orthonormalized coupling coefficients. 

\vskip.2cm

In addition to a simple resolution of the outer multiplicity issue above, the method presented herein is also advantageous regarding the choice of the phase factor of the CGC's. As the solution of the null space is given in terms of column vectors, one only needs to ensure that the first non-vanishing entry belonging to each vector is always positive, and then the phase of the rest in the column vector simply follows. It will be shown in the next subsection that when the signs of the CGC's associated with the highest weight state of the coupled irrep are fixed, the signs of the CGC's belonging to other states (including the lowest weight state) are determined by the actions of the lowering generators.

\subsection{U(4)$\supset$U(3)$\supset$U(2)$\supset$U(1) CGC's for the Lower Weight States}
\label{secIVD}

On the grounds that the action of the lowering operators $E_{43}''$, $E_{32}''$, and $E_{21}''$, see Eq. (\ref{eq:lowering}) and Appendix \ref{appendA}, are known, one might assume that it is straightforward to go through to all other states from the highest weight one via those generators. Nonetheless, a closer look at the action thereof on any state $\ket{G''}_\eta$ (including $\ket{HW''}_\eta$) reveals a challenge owing to the fact that there are up to three different Gelfand states which can result from applying the lowering generators. Hence, it is more practical to evaluate CGC's for all states of the same p-weight $[w_4'',w_3'',w_2'',w_1'']$ at the same time. 

\vskip.2cm

Suppose $\ket{G''}_\eta$ is one of those states, and a Gelfand state which under the action of either $E_{43}''$ , $E_{32}''$, or $E_{21}''$, returns to $\ket{G''}_\eta$ is called a parent state of $\ket{G''}_\eta$. There are six such states, which can be expanded in the uncoupled basis as follows,
\begin{small}
\begin{equation}
\begin{array}{c}
    \ket{G''+I_{11}}_\eta = \sum_{G,G'} C_{G,G'}^{G''_\eta+I_{11}} \ket{G}\ket{G'}, 
    \ket{G''+I_{12}}_\eta = \sum_{G,G'} C_{G,G'}^{G''_\eta+I_{12}} \ket{G}\ket{G'}, 
    \ket{G''+I_{22}}_\eta = \sum_{G,G'} C_{G,G'}^{G''_\eta+I_{22}} \ket{G}\ket{G'}, \\
    \ket{G''+I_{13}}_\eta = \sum_{G,G'} C_{G,G'}^{G''_\eta+I_{13}} \ket{G}\ket{G'},
    \ket{G''+I_{23}}_\eta = \sum_{G,G'} C_{G,G'}^{G''_\eta+I_{23}} \ket{G}\ket{G'},
    \ket{G''+I_{33}}_\eta = \sum_{G,G'} C_{G,G'}^{G''_\eta+I_{33}} \ket{G}\ket{G'},
\end{array}
\end{equation}
\end{small}
where the expansion coefficients are presumably known beforehand. By acting $E_{i+1,i}''=E_{i+1,i}\otimes \mathbf{1}' + \mathbf{1}\otimes E_{i+1,i}'$ (with $i=1,2,3$) on these equations, one obtains
\begin{align}
    e_{21}(G''+I_{11})\ket{G''}_\eta = \sum_{G,G'}\, C_{G,G'}^{G''_\eta+I_{11}} \bigg[e_{21}(G)\ket{G-I_{11}}\ket{G'} + e_{21}(G')\ket{G}\ket{G'-I_{11}} \bigg], 
\end{align}
\begin{align}
   &  e_{32,12}(G''+I_{12})\ket{G''}_\eta + e_{32,22}(G''+I_{12})\ket{G''+I_{12}-I_{22}}_\eta = \nonumber\\
   &\sum_{G,G'}\, C_{G,G'}^{G''_\eta+I_{12}}\, \bigg[e_{32,12}(G)\ket{G-I_{12}}\ket{G'} + e_{32,22}(G)\ket{G-I_{22}}\ket{G'}  \nonumber \\
    &+ e_{32,12}(G')\ket{G}\ket{G'-I_{12}} + e_{32,22}(G')\ket{G}\ket{G'-I_{22}} \bigg],
\end{align}
\begin{align}
    & e_{32,12}(G''+I_{22})\ket{G''+I_{22}-I_{12}}_\eta + e_{32,22}(G''+I_{22})\ket{G''}_\eta = \nonumber\\
    &\sum_{G,G'}\, C_{G,G'}^{G''_\eta+I_{22}} \bigg[e_{32,12}(G)\ket{G-I_{12}}\ket{G'} + e_{32,22}(G)\ket{G-I_{22}}\ket{G'} \nonumber\\
    &+ e_{32,12}(G')\ket{G}\ket{G'-I_{12}} + e_{32,22}(G')\ket{G}\ket{G'-I_{22}} \bigg],
\end{align}
\begin{align}
    & e_{43,13}(G''+I_{13})\ket{G''}_\eta + e_{43,23}(G''+I_{13})\ket{G''+I_{13}-I_{23}}_\eta + e_{43,33}(G''+I_{13})\ket{G''+I_{13}-I_{33}}_\eta = \nonumber\\
    &\sum_{G,G'}\, C_{G,G'}^{G''_\eta+I_{13}} \bigg[e_{43,13}(G)\ket{G-I_{13}}\ket{G'} + e_{43,23}(G)\ket{G-I_{23}}\ket{G'} + e_{43,33}(G)\ket{G-I_{33}}\ket{G'} \nonumber\\
    &+ e_{43,13}(G')\ket{G}\ket{G'-I_{13}} + e_{43,23}(G')\ket{G}\ket{G'-I_{23}} + e_{43,33}(G')\ket{G}\ket{G'-I_{33}} \bigg], 
\end{align}
\begin{align}
    & e_{43,13}(G''+I_{23})\ket{G''+I_{23}-I_{13}}_\eta + e_{43,23}(G''+I_{23})\ket{G''}_\eta + e_{43,33}(G''+I_{23})\ket{G''+I_{23}-I_{33}}_\eta = \nonumber\\
    &\sum_{G,G'}\, C_{G,G'}^{G''_\eta+I_{23}} \bigg[e_{43,13}(G)\ket{G-I_{13}}\ket{G'} + e_{43,23}(G)\ket{G-I_{23}}\ket{G'} + e_{43,33}(G)\ket{G-I_{33}}\ket{G'} \nonumber\\
    &+ e_{43,13}(G')\ket{G}\ket{G'-I_{13}} + e_{43,23}(G')\ket{G}\ket{G'-I_{23}} + e_{43,33}(G')\ket{G}\ket{G'-I_{33}} \bigg],
\end{align}
\begin{align}
    & e_{43,13}(G''+I_{33})\ket{G''+I_{33}-I_{13}}_\eta + e_{43,23}(G''+I_{33})\ket{G''+I_{33}-I_{23}}_\eta + e_{43,33}(G''+I_{33})\ket{G''}_\eta = \nonumber\\
    &\sum_{G,G'}\, C_{G,G'}^{G''_\eta+I_{33}} \bigg[e_{43,13}(G)\ket{G-I_{13}}\ket{G'} + e_{43,23}(G)\ket{G-I_{23}}\ket{G'} + e_{43,33}(G)\ket{G-I_{33}}\ket{G'} \nonumber\\
    &+ e_{43,13}(G')\ket{G}\ket{G'-I_{13}} + e_{43,23}(G')\ket{G}\ket{G'-I_{23}} + e_{43,33}(G')\ket{G}\ket{G'-I_{33}} \bigg].
\end{align}
It is apparent in these equations that the states $\ket{G''}_\eta$, $\ket{G''+I_{12}-I_{22}}_\eta$, $\ket{G''+I_{22}-I_{12}}_\eta$, $\ket{G''+I_{13}-I_{23}}_\eta$, $\ket{G''+I_{13}-I_{33}}_\eta$, $\ket{G''+I_{23}-I_{13}}_\eta$, $\ket{G''+I_{23}-I_{33}}_\eta$, $\ket{G''+I_{33}-I_{13}}_\eta$, and $\ket{G''+I_{33}-I_{23}}_\eta$, have the same p-weight -- which is the justification for computing the set of CGC's for all states having the same p-weight simultaneously. Since the expansions of the parent states in the uncoupled basis are assumed to be known, one obtains all the coefficients $C_{G,G'}^{G''_\eta}$ by arranging these equations into a matrix equation and then inverting the matrix on the left hand side which subsumes the coefficients resulted from acting the lowering generators on $\ket{G''+I_{11}}_\eta, \ket{G''+I_{12}}_\eta, \ket{G''+I_{22}}_\eta,
    \ket{G''+I_{13}}_\eta, \ket{G''+I_{23}}_\eta, \ket{G''+I_{33}}_\eta$,  as matrix elements. It is worth noting that as these equations are in general not linearly independent  \cite{Alex2011JMP}, one should eliminate those equations that depends on one another to avoid the solution being overdetermined.

\section{U(4) Racah Recoupling Coefficients}
\label{secV}

Similarly to the SU(2) case, where the 6-$j$ and 9-$j$ symbols are needed when more than two angular momenta are coupled together, the coupling of more than two U(4) irreps also necessitates the so-called Racah recoupling coefficients, which by definition provide a transformation between different orders of coupling. In this section, the formulae to compute the recoupling coefficients for three and four U(4) irreps are presented; moreover, as a matter of convention, we call them 6-U(4) and 9-U(4) coefficients to be consistent with the naming in the SU(2) case. Also, for convenience, in what follows, we use a shorthand notation for U(4) irreps: $f \equiv [n_{14},n_{24},n_{34},n_{44}]$. 

\subsection{Coupling of Three U(4) Irreps: 6-U(4) Recoupling Coefficients} 
\begin{figure}[h!]
    \centering
    \includegraphics[scale=0.5]{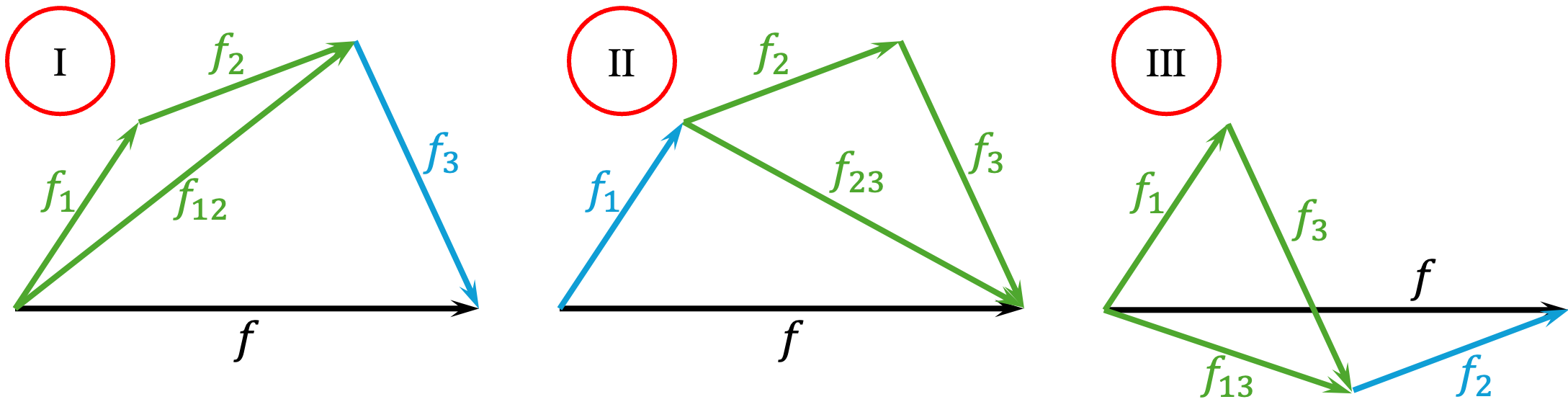}
    \caption{Three different orders of coupling three U(4) irreps  $f_1$, $f_2$ and $f_3$.}
    \label{fig:6-U(4)}
\end{figure}

There are three distinct orders to couple three U(4) irreps $f_1$, $f_2$ and $f_3$ to a final irrep $f$, namely
\begin{align}
    (f_1 \otimes f_2) \otimes f_3, \text{  } f_1 \otimes (f_2 \otimes f_3), \text{  and  } (f_1 \otimes f_3) \otimes f_2,
\end{align}
as shown in FIG. \ref{fig:6-U(4)}. Similarly to the U(3) case \cite{Dang2024}, there are two types of recoupling coefficients, one of which involves a permutation of U(4) irreps. The first type is referred to as the $U$-coefficients, which give the transformation between the first two schemes (I and II) and can be calculated by using the following relation
\begin{align}
    \sum_{\eta_{1,23}} C_{G_1,G_{23}}^{G_{\eta_{1,23}}} \times U(f_1 f_2 f f_3;f_{12}\eta_{12}\eta_{12,3},f_{23}\eta_{23}\eta_{1,23}) = \sum_{G_2 G_3 G_{12}} C_{G_1,G_2}^{G_{12,\eta_{12}}} \times C_{G_{12},G_3}^{G_{\eta_{12,3}}} \times C_{G_2,G_3}^{G_{23,\eta_{23}}}.
\end{align}
As the CGC's on the right hand side are generated for the highest weight states first (see Section \ref{secIV}), one can choose $\ket{G_{23}}=\ket{HW_{23}}$ and $\ket{G}=\ket{HW}$, while letting $\ket{G_1}$ run over its range, to obtain a set of linear equations where the $U$-coefficients are the unknowns. It follows that all the $U$-coefficients for $\eta_{1,23}=1,2,...,\eta_{1,23}^{\max}$ can be evaluated simultaneously by solving this set of linear equations. 

\vskip.2cm

The second type of 6-U(4) coefficients is the so-called $Z$-coefficients that provide the transformation between schemes I and III. Analogously to the $U$-coefficients, by fixing $\ket{G_{13}}=\ket{HW_{13}}$ and $\ket{G}=\ket{HW}$, while letting $\ket{G_2}$ run over its range in the relation
\begin{align}
    \sum_{\eta_{13,2}} C_{G_{13},G_2}^{G_{\eta_{13,2}}} \times Z(f_2 f_1 f f_3;f_{12}\eta_{12}\eta_{12,3},f_{13}\eta_{13}\eta_{13,2}) = \sum_{G_1 G_3 G_{12}} C_{G_1,G_2}^{G_{12,\eta_{12}}} \times C_{G_{12},G_3}^{G_{\eta_{12,3}}} \times C_{G_1,G_3}^{G_{13,\eta_{13}}},
    \label{eq:Z6}
\end{align}
one arrives at a set of linear equations that can be solved to obtain all the $Z$-coefficients for $\eta_{13,2}=1,2,...,\eta_{13,2}^{\max}$ simultaneously.

\subsection{Coupling of Four U(4) Irreps: 9-U(4) Recoupling Coefficients}

It should be cleart that the coupling of four U(4) irreps $f_1$, $f_2$, $f_3$ and $f_4$ to a final one $f$, requires higher complexity than the previous case of three U(4) irreps as there are more orders of coupling. However, once the 6-U(4) coefficients are calculated based on the above description, they can be utilized to evaluate the so-called 9-U(4) coefficients, which give the transformation between different schemes. For the two orders of coupling, $(f_1 \otimes f_2) \otimes (f_3 \otimes f_4)$ and $(f_1 \otimes f_3) \otimes (f_2 \otimes f_4)$, the recoupling coefficients are computed via the following formula:
\begin{multline}
    \Biggl\{ \begin{array}{cccc}
         f_1& f_2& f_{12}& \eta_{12} \\
         f_3& f_4& f_{34}& \eta_{34} \\
         f_{13}& f_{24}& f& \eta_{13,24} \\
         \eta_{13}& \eta_{24}& \eta_{12,34}
    \end{array} \Biggl\} = \sum_{f_0, \eta_{13,2} \eta_{04} \eta_{12,3}} U(f_{13} f_2 f f_4;f_{0}\eta_{13,2}\eta_{04},f_{24}\eta_{24}\eta_{13,24})  \times \\ Z(f_2 f_1 f_0 f_3;f_{12}\eta_{12}\eta_{12,3},f_{13}\eta_{13}\eta_{13,2})
    \times U(f_{12} f_3 f f_4;f_{0}\eta_{12,3}\eta_{04},f_{34}\eta_{34}\eta_{12,34}),
\end{multline}
where the $f_0$'s are resultant of coupling three U(4) irreps $f_1$, $f_2$ and $f_3$, in both schemes: $(f_1 \otimes f_2) \otimes f_3$ and $(f_1 \otimes f_3) \otimes f_2$.

\section{Application in Nuclear Physics -- U(4)$\supset$SU$_S$(2)$\otimes$SU$_T$(2) Wigner Coefficients}
\label{secVI}

\subsection{Physical Reduction in Nuclear Physics}
\label{secVIA}

In Wigner's Supermultiplet theory, the relevant ``physical" group-subgroup chain is the following:

\begin{equation}
    \begin{array}{cccccccc}
        {\rm U(4)} &\supset &{\rm SU_S(2)} &\otimes &{\rm SU_T(2)} \supset &{\rm O_S(2)} &\otimes &{\rm O_T(2)}  \\
        |[n_{14},n_{24},n_{34},n_{44}] &\zeta &{S} & &{T;} &{M_S} & &{M_T}\rangle 
    \end{array} 
\end{equation}
wherein the label $\zeta$ is the inner multiplicity accounting for multiple occurrences of a $(S,T)$ pair within a given U(4) irrep $[n_{14},n_{24},n_{34},n_{44}]$. The physical generators associated with the quantum numbers $S,T,M_S,M_T$ can be expressed in terms of the canonical generators, Eqs. (\ref{eq:remain}), (\ref{eq:raising}), (\ref{eq:lowering}), as follows \cite{Pan2023EPJP}:
\begin{equation}
    \begin{array}{ccc}
         S_+ = E_{13}+E_{24}, & S_0 = (E_{11}+E_{22}-E_{33}-E_{44})/2, & S_- = (E_{31}+E_{42})  \\
         T_+ = E_{12}+E_{34}, & T_0 = (E_{11}-E_{22}+E_{33}-E_{44})/2, & T_- = (E_{21}+E_{43}) 
    \end{array}
\end{equation}
\begin{figure}[h!]
    \centering
    \includegraphics[scale=0.6]{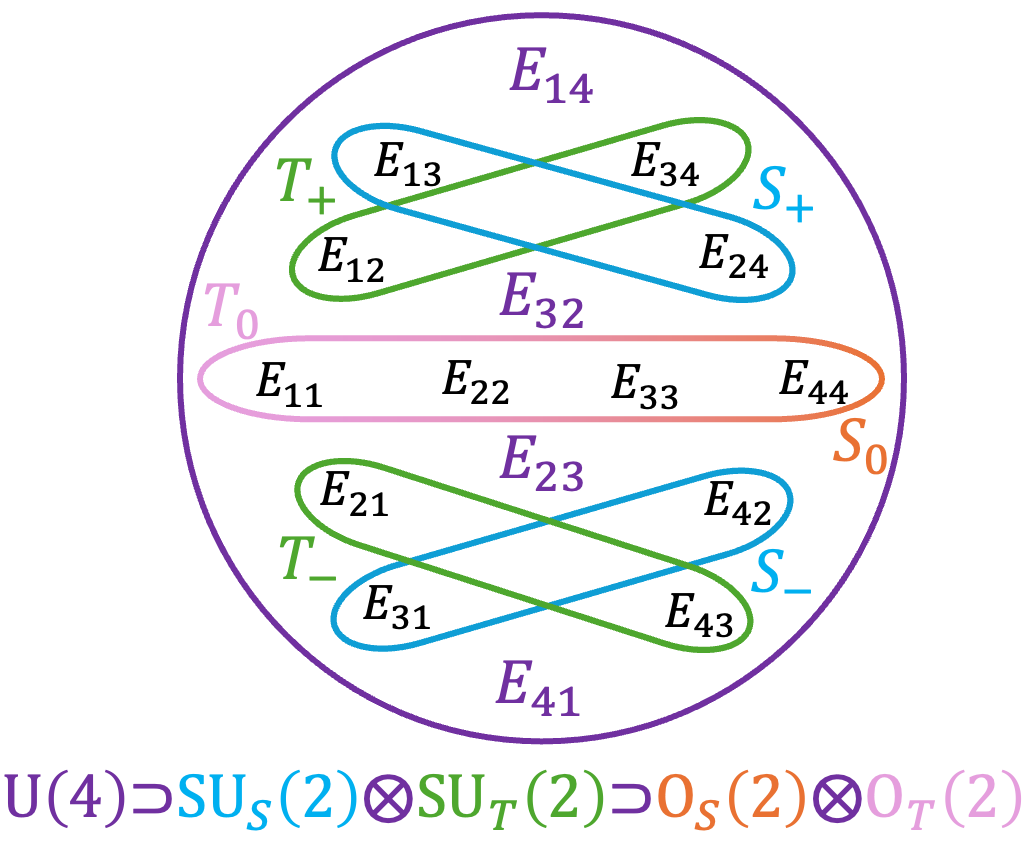}
    \caption{The group-subgroup chain realization of U(4) in nuclear physics.}
    \label{fig:phys-chain}
\end{figure}

\begin{figure}[h!]
    \centering
    \includegraphics[scale=0.7]{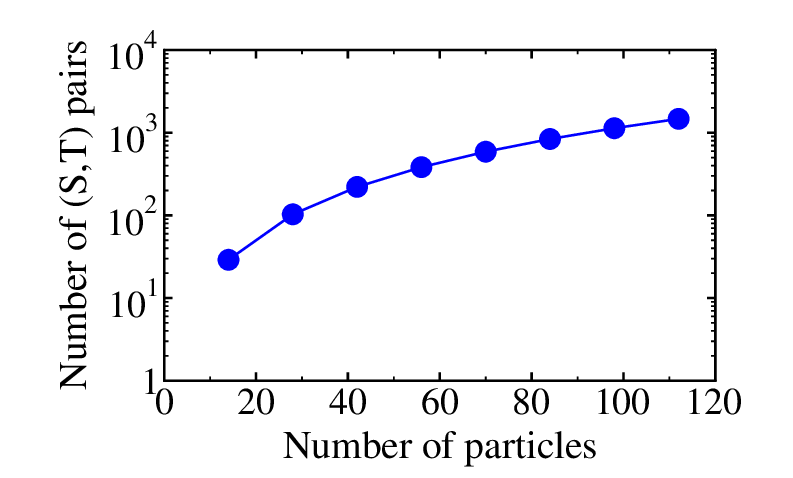}
    \caption{The number of $(S,T)$ pairs that are allowed in the reduction $\rm U_{ST}(4)$ for some U(4) irreps of the type $[4k,2k,k,0]$ (with even $k$), where $A= \sum_i n_{i4} = 7k$ is the number of nucleons that are distributed into the irreps.}
    \label{fig:STpair}
\end{figure}

Figure \ref{fig:phys-chain} is a graphical demonstration of the group-subgroup structure of the physical chain $\rm U_{ST}(4)$. On the grounds that O(6) and SU(4) are locally isomorphic and that a U(4) irrep $[n_{14},n_{24},n_{34},n_{44}]$ and a SU(4) irrep $[n_{14}-n_{44},n_{24}-n_{44},n_{34}-n_{44}]$ are equivalent, the allowed values of the spin and isospin quantum numbers, $S$ and $T$, can be determined from the O(6) quantum numbers
\begin{equation}
    p_1=\frac{1}{2}(n_{14}+n_{24}-n_{34}-n_{44}), p_2=\frac{1}{2}(n_{14}-n_{24}+n_{34}-n_{44}) \text{ and } p_3=\frac{1}{2}(n_{14}-n_{24}-n_{34}+n_{44})
\end{equation}
via the following constrain \cite{Partensky1978JMP,Pan2023EPJP}:
\begin{equation}
    \Gamma_{\min} \le S,T \le p_1 \text{ and } \Gamma_{\min} \le S+T \le p_1+p_2 \text{, where } \Gamma_{\min} = 
    \begin{cases}
        0 \text{ if } 2p_1 \text{ is even}\\
        1/2 \text{ otherwise}
    \end{cases},
\end{equation}
which leads to an increasing number of valid $(S,T)$ pairs as the number of particles grows, see FIG. \ref{fig:STpair}.

\begin{figure}[h!]
    \centering
    \includegraphics[scale=0.4]{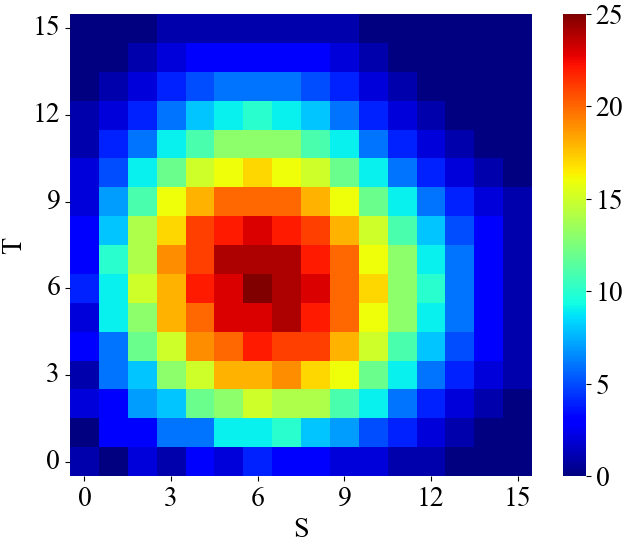}
    \caption{The inner multiplicity $\zeta_{\max}$ of spin ($S$) and isospin ($T$) quantum numbers for the U(4) irrep $[64,32,16,0]$.}
    \label{fig:multi}
\end{figure}

Racah \cite{Racah1949RMP} was the first to provide a resolution for the inner multiplicity of ($S,T$) pairs within a U(4) irrep by contracting a general tensor to a symmetric tensor of character $[f,0,0,0]$. The formula can be summarized as follows: for a given U(4) irrep $[n_{14},n_{24},n_{34},n_{44}]$, the inner multiplicity of a particular $(S,T)$ pair is 
\begin{equation}
    \zeta_{\max}(S,T) = \Omega(n_{14}-n_{34},n_{24}-n_{44},S,T) - \Omega(n_{14}-n_{44}+1,n_{24}-n_{34}-1,S,T) - \Omega(n_{14}-n_{24}-1,n_{34}-n_{44}-1,S,T),
\end{equation}
where the $\Omega$ function is nonzero only if $(x+y)/2 \ge S,T$, in which case it is defined as
\begin{equation}
    \Omega(x,y,S,T) = \phi(y+2-|T-S|) - \phi(y+1-T-S) + \phi(T+S-x-1) - \phi(T+S-|T-S|-x+y+1)/2,  
\end{equation}
for $x \ge y$, otherwise $\Omega(x,y,S,T)=\Omega(y,x,S,T)$. Here $\phi(x) = \text{Int}[x^2/4]$ if $x > 0$ and vanishes otherwise. Figure \ref{fig:multi} illustrates an example of how the inner multiplicity changes with spin and isospin quantum numbers for the U(4) irrep $[64,32,16,0]$, and it can be seen that high multiplicities are found in middle range of spin and isospin values, which is an expected feature of atomic nuclei.

\vskip.2cm

On the grounds that the atomic nucleus itself is a composite system with nucleons or clusters thereof as constituents and that each of those possesses a $\rm U_{ST}(4)$ symmetry, it is important know the coupling basis states in the physical scheme. Since the recoupling coefficients are independent of the subgroup chains, the formulas presented in Section \ref{secV} still hold and do not need to be reviewed. The Wigner $\rm U_{ST}(4)$ coupling coefficients are, on the contrary, different; however, they can be computed via the canonical coupling coefficients from Section \ref{secIV}. In the following subsections, we first present the transformation between the physical and canonical basis states, and then the coupling coefficients in the physical basis using the transformation.

\subsection{Transformation between the Physical and the Canonical Bases}
\label{secVIB}

The $\rm U_{ST}(4)$ basis states were constructed from the canonical $\rm U(4) \supset U(3) \supset U(2) \supset U(1)$ ones by integrating Wigner $D$-matrices over solid angles by Draayer in 1970 \cite{Draayer1970JMP}. However, this procedure is somewhat cumbersome and difficult for computer implementation. Therefore, in this article, we demonstrate a modern projection method presented recently in \cite{Pan2023EPJP} based on the null space concept of the spin and isospin generators, which is very distinct from the projection technique employed by Draayer.

\vskip.2cm

Since there is always a freedom for the starting point when computing Wigner coefficients -- i.e., one can either construct from the highest weight state ($M_S=S$ and $M_T=T$) or the lowest weight state ($M_S=-S$ and $M_T=-T$), and then generate all other states by applying consecutively the lowering ($S_-$ and $T_-$) or raising ($S_+$ and $T_+$) operators -- here, we follow the standard procedure that is widely used to evaluate SU(2) CGC's and start with the highest weight state
\begin{align*}
    \ket{\begin{array}{c}
n_{14},n_{24},n_{34},n_{44}\cr
\zeta S M_S=S, T M_T=T
\end{array}},
\end{align*}
and once it is known, other states can be obtained by applying the lowering operator as follows,
\begin{align}
    \ket{\begin{array}{c}
n_{14},n_{24},n_{34},n_{44}\cr
\zeta S M_S, T M_T
\end{array}} = \sqrt{\frac{(S+M_S)!(T+M_T)!}{(2S)!(S-M_S)!(2T)!(T-M_T)!}} (S_-)^{S-M_S} (T_-)^{T-M_T} \ket{\begin{array}{c}
n_{14},n_{24},n_{34},n_{44}\cr
\zeta S M_S=S, T M_T=T
\end{array}}, 
\label{eq:lower-weight}
\end{align}
where $S_- = E_{42}+E_{31}$ and $T_-=E_{43}+E_{21}$.

\vskip.2cm

For each $(S,T)$ pair, the highest weight basis state can be expanded in the canonical basis as follows \cite{Pan2023EPJP},
\begin{small}
\begin{align}
    \ket{\begin{array}{c}
n_{14},n_{24},n_{34},n_{44}\cr
\zeta S M_S=S, T M_T=T
\end{array}} = \sum_{a_1 a_2 a_3 q} C^{\zeta}_{a_1 a_2 a_3 q}([n_{14},n_{24},n_{34},n_{44}]SM_S=S,TM_T=T) \Bigg| \begin{array}{c}
n_{14},n_{24},n_{34},n_{44}\cr
n_{14}-a_1,n_{24}-a_2,n_{34}-a_3\cr
n_{14}-a_1-q, n/2+S-n_{14}+a_1+q\cr
n_{44}+S+T+a_1+a_2+a_3
\end{array} \Bigg\rangle,
\label{eq:expansion}
\end{align}
\end{small}
where the parameters are
\begin{align}
    n=\sum_{i=1}^4 n_{i4}, 0 \le a_1 \le n_{14}-n_{24}, 0 \le a_2 \le n_{24}-n_{34}, 0 \le a_3 \le n_{34}-n_{44} \text{ and } q_{\min}(S) \le q \le q_{\max}(S,T)
\end{align}
with the boundaries of $q$ defined as
\begin{equation}
    q_{\min}(S) = \max(0,n_{14}+n_{34}-n/2-S-a_1-a_3)
\end{equation}
and 
\begin{multline}
    q_{\max}(S,T) = \min(n_{14}+n_{24}-n/2-S-a_1-a_2,n_{14}-n_{44}-S-T-2a_1-a_2-a_3, \\ 
    n_{14}+n_{44}-n/2+T+a_2+a_3,n_{14}-n_{24}-a_1+a_2).
\end{multline}

The expansion coefficients $C^{\zeta}_{a_1 a_2 a_3 q}([n_{14},n_{24},n_{34},n_{44}]SM_S=S,TM_T=T)$ can be found using the fact that the highest weight state, by definition, must vanish under the action of the raising operators $S_+$ and $T_+$:
\begin{align}
    S_+ \ket{\begin{array}{c}
n_{14},n_{24},n_{34},n_{44}\cr
\zeta SM_S=S, TM_T=T
\end{array}} = (E_{13}+E_{24})\ket{\begin{array}{c}
n_{14},n_{24},n_{34},n_{44}\cr
\zeta SM_S=S, TM_T=T
\end{array}} &= 0, \\
T_+ \ket{\begin{array}{c}
n_{14},n_{24},n_{34},n_{44}\cr
\zeta SM_S=S, TM_T=T
\end{array}} = (E_{12}+E_{34})\ket{\begin{array}{c}
n_{14},n_{24},n_{34},n_{44}\cr
\zeta SM_S=S, TM_T=T
\end{array}} &= 0.   
\end{align}
Plugging into Eq. (\ref{eq:expansion}), one can obtain
\begin{small}
\begin{align}
 \sum_{a_1 a_2 a_3 q} C^{\zeta}_{a_1 a_2 a_3 q}([n_{14},n_{24},n_{34},n_{44}]SM_S=S,TM_T=T) (E_{13}+E_{24}) \Bigg| \begin{array}{c}
n_{14},n_{24},n_{34},n_{44}\cr
n_{14}-a_1,n_{24}-a_2,n_{34}-a_3\cr
n_{14}-a_1-q, n/2+S-n_{14}+a_1+q\cr
n_{44}+S+T+a_1+a_2+a_3
\end{array} \Bigg\rangle = 0, \\
\sum_{a_1 a_2 a_3 q} C^{\zeta}_{a_1 a_2 a_3 q}([n_{14},n_{24},n_{34},n_{44}]SM_S=S,TM_T=T) (E_{12}+E_{34}) \Bigg| \begin{array}{c}
n_{14},n_{24},n_{34},n_{44}\cr
n_{14}-a_1,n_{24}-a_2,n_{34}-a_3\cr
n_{14}-a_1-q, n/2+S-n_{14}+a_1+q\cr
n_{44}+S+T+a_1+a_2+a_3
\end{array}\Bigg\rangle = 0.
\end{align}
\end{small}

The action of the generators $E_{12}$, $E_{13}$, $E_{24}$ and $E_{34}$ on the canonical basis states as given in Section \ref{secIII} results in a set of homogeneous linear equations wherein the expansion coefficients $C^{\zeta}_{a_1 a_2 a_3 q}([n_{14},n_{24},n_{34},n_{44}]SM_S=S,TM_T=T)$ are the unknowns. This set of linear equations can be written in a compact matrix form
\begin{align}
    \mathbf{P}([n_{14},n_{24},n_{34},n_{44}],ST) \mathbf{C}^{\zeta} = \mathbf{0},
\end{align}
where the matrix $\mathbf{P}([n_{14},n_{24},n_{34},n_{44}],ST)$ is called the spin-isospin projection matrix and the entries of the column vectors $\mathbf{C}^{\zeta}$ for $\zeta=1,2,...,\zeta_{\max}$ are the expansion coefficients that provide the transformation between the physical and canonical bases. It follows that finding the transformation is tantamount to solving the null space of the spin-isospin projection matrix; and all expansion coefficients associated with different $\zeta$'s are obtained at the same time. Then the inner multiplicity that occurs in the $\rm U_{ST}(4)$ reduction is deduced by checking the dimension of the null space of $\mathbf{P}([n_{14},n_{24},n_{34},n_{44}],ST)$. To be more specific, if the null space is found to be empty, the $(S,T)$ pair under consideration is not allowed for the given U(4) irrep; if there is a single and unique solution, the inner multiplicity of $(S,T)$ pair is $\zeta_{\max}=1$. It happens more often that the solution to the null space, hence the expansion coefficients, are not unique and somewhat arbitrary, in which case, the $(S,T)$ pair is degenerate within the U(4) irrep of interest. It is also worth noting that, as pointed out eariler in in Section \ref{secIVC}, solving the null space of a matrix, especially if done numerically, often does not return orthonormalized column vectors, in which case a Gram-Schmidt orthonormalization can be carried out as it is done for $\mathbf{P}(HW'')$. Concerning the phase of the expansion coefficients from the physical to the canonical basis, one only needs to fix the first non-zero element of each column vector $\mathbf{C}^\zeta$ to be positive, and then the phase of the other elements simply follows.

\vskip.2cm

A quick dimensionality check can be carried out to validate whether the solution of the null space of all the spin-isospin projection matrices is complete, especially when it is done by numerical methods, via the relation
\begin{align}
    \dim([n_{14},n_{24},n_{34},n_{44}]) = \sum_{S,T} \zeta_{\max}(S,T) (2S+1)(2T+1),
\end{align}
since $M_S$ takes values $S,S-1,...,-S+1,-S$ and similarly for $M_T$.

\vskip.2cm

Once the coefficients associated with the construction of the physical highest weight state in terms of Gelfand states, one can calculate the expansion of all other states using Eq. (\ref{eq:lower-weight}) as follows
\begin{small}
\begin{align}
     \ket{\begin{array}{c}
n_{14},n_{24},n_{34},n_{44}\cr
\zeta S M_S=S-\alpha; T M_T=T-\beta
\end{array}} &= \sqrt{\frac{(2S-\alpha)!(2T-\beta)!}{(2S)!\alpha!(2T)!\beta!}}  \sum_{a_1 a_2 a_3 q} C^{\zeta}_{a_1 a_2 a_3 q}([n_{14},n_{24},n_{34},n_{44}]SM_S=S,TM_T=T) \nonumber \\ 
    & \times (E_{42}+E_{31})^{\alpha} (E_{43}+E_{21})^{\beta} \Bigg| \begin{array}{c}
n_{14},n_{24},n_{34},n_{44}\cr
n_{14}-a_1,n_{24}-a_2,n_{34}-a_3\cr
n_{14}-a_1-q, n/2+S-n_{14}+a_1+q\cr
n_{44}+S+T+a_1+a_2+a_3
\end{array}\Bigg\rangle \nonumber \\
&= \sqrt{\frac{(2S-\alpha)!(2T-\beta)!}{(2S)!\alpha!(2T)!\beta!}} \sum_G C_{a_1 a_2 a_3 q}^\zeta ([n_{14},n_{24},n_{34},n_{44}]SM_S=S-\alpha,TM_T=T-\beta)\ket{G},
\end{align}
\end{small}
where $\alpha=S-M_S=0,1,...,2S$, $\beta=T-M_T=0,1,...,2T$ and $\ket{G}$'s are Gelfand states that occur as a result of the action of the generators $E_{42}$, $E_{31}$, $E_{43}$ and $E_{21}$ on the Gelfand states involved in the expansion of the highest weight state. Since defining the action of the U(4) generators as methods associated with a data structure is feasible in object-oriented programming languages like C++, we neither give an explicit expression here for the expansion coefficients $C_{a_1 a_2 a_3 q}^\zeta ([n_{14},n_{24},n_{34},n_{44}]SM_S=S-\alpha,TM_T=T-\beta)$ nor use it in our code. Nonetheless, for the matter of interest, the previous work \cite{Pan2023NPA} contains explicit formulae for these coefficients in Eqs. (11)-(15).

\subsection{U(4)$\supset$SU$_S$(2)$\otimes$SU$_T$(2) Wigner Coefficients}
\label{secVIC}

On the grounds that any Wigner coefficient $\rm U(4)\supset SU_S(2) \otimes SU_T(2) \supset O_S(2) \otimes O_T(2) $, 
\begin{multline}
    \braket{\begin{array}{c}
n_{14},n_{24},n_{34},n_{44}\cr
\zeta S M_S, T M_T
\end{array}; \begin{array}{c}
n_{14}',n_{24}',n_{34}',n_{44}'\cr
\zeta' S' M_S', T' M_T'
\end{array}}{\begin{array}{c}
n_{14}'',n_{24}'',n_{34}'',n_{44}''\cr
\zeta'' S'' M_S'', T'' M_T''
\end{array}}_\eta = \braket{SM_S;S'M_S'}{S''M_S''} \times \braket{TM_T;T'M_T'}{T''M_T''} \times \\
\bra{\begin{array}{c}
n_{14},n_{24},n_{34},n_{44}\cr
\zeta S, T 
\end{array}; \begin{array}{c}
n_{14}',n_{24}',n_{34}',n_{44}'\cr
\zeta' S' , T' 
\end{array}}\text{} \ket{\begin{array}{c}
n_{14}'',n_{24}'',n_{34}'',n_{44}''\cr
\zeta'' S'' , T'' 
\end{array}}_\eta.
\end{multline}
can be factorized into two usual $\rm SU(2)$ CGC's, which are well-known and can be evaluated using various existing computer programs, and a reduced (or double-barred) coefficient, this paper focuses on the computation of the reduced coefficients. 

\vskip.2cm

Pan et al. \cite{Pan2023NPA} provided a formula for the evaluation of $\rm U_{ST}(4)$ Wigner coefficients by taking advantage of $\rm U(4) \supset U(3) \supset U(2) \supset U(1)$ CGC's and the transformation between the two physical and canonical bases, which is also employed in this paper, however, without the factorization of the canonical CGC's into $\rm U(4) \supset U(3)$, $\rm U(3)\supset U(2)$ and $\rm U(2)\supset U(1)$ pieces as follows,
\begin{scriptsize}
\begin{multline}
\bra{\begin{array}{c}
n_{14},n_{24},n_{34},n_{44}\cr
\zeta S, T 
\end{array}; \begin{array}{c}
n_{14}',n_{24}',n_{34}',n_{44}'\cr
\zeta' S' , T' 
\end{array}}\text{}\ket{\begin{array}{c}
n_{14}'',n_{24}'',n_{34}'',n_{44}''\cr
\zeta'' S'' , T'' 
\end{array}}_\eta  =  \braket{SS;S'S'-\alpha}{S''S''}^{-1} \braket{TT;T'T'-\beta}{T''T''}^{-1}  \\
\times \sqrt{\frac{(2S'-\alpha)!(2T'-\beta)!}{(2S')!\alpha!(2T')!\beta!}} \sum_{a_i q,a_i'q',a_i''q''} C_{a_1 a_2 a_3 q}^{\zeta}([n_{14},n_{24},n_{34},n_{44}]SM_S=S,TM_T=T) \\ \times C_{a_1'a_2'a_3'q'}^{\zeta'}([n_{14}',n_{24}',n_{34}',n_{44}']S'M_S'=S'-\alpha,T'M_T'=T'-\beta)  C_{a_1'' a_2'' a_3'' q''}^{\zeta''}([n_{14}'',n_{24}'',n_{34}'',n_{44}'']S''M_S''=S'',T''M_T''=T'') \\
\times 
\Bigg\langle \begin{array}{c}
n_{14},n_{24},n_{34},n_{44}\cr
n_{14}-a_1,n_{24}-a_2,n_{34}-a_3\cr
n_{14}-a_1-q, n/2+S-n_{14}+a_1+q\cr
n_{44}+S+T+a_1+a_2+a_3
\end{array} ; \begin{array}{c}
n_{14}',n_{24}',n_{34}',n_{44}'\cr
n_{14}'-a_1',n_{24}'-a_2',n_{34}'-a_3'\cr
n_{14}'-a_1'-q', n'/2+S'-n_{14}'+a_1'+q'\cr
n_{44}'+S'+T'+a_1'+a_2'+a_3'
\end{array} \Bigg| \begin{array}{c}
n_{14}'',n_{24}'',n_{34}'',n_{44}''\cr
n_{14}''-a_1'',n_{24}''-a_2'',n_{34}''-a_3''\cr
n_{14}''-a_1''-q'', n''/2+S''-n_{14}''+a_1''+q''\cr
n_{44}''+S''+T''+a_1''+a_2''+a_3''
\end{array} \Bigg\rangle,
\end{multline}
\end{scriptsize}
where $\alpha=S+S'-S''=0,1,...,2S'$ and $\beta=T+T'-T''=0,1,...,2T'$. There are two reasons for not using the factorization: i) the entire coefficients are known using the method described in Section \ref{secIV}, and ii) one would suffer from another outer multiplicity issue owing to the coupling of two U(3) irreps. 

Tables \ref{tab:eg1} and \ref{tab:eg2} provides two simple examples for Wigner coefficients computed from our C++ library using the methodology presented in this article. The former shows the elementary Wigner coefficients \cite{Pan2024CPC}, meaning that one of the irrep that is coupled is $[1,0,0,0]$, in which case, the coupling is always free of the outer multiplicity and the reduction to $\rm SU_S(2)\otimes SU_T(2)$ is also unique. Meanwhile, the latter shows a coupling of two irreps $[8,4,2,0] \otimes [1,1,0,0]$, which is quite nontrivial and more complicated. It can be verified easily that the following orthogonality relations for the reduced Wigner coefficients
\begin{multline}
    \sum_{\zeta,S,T,\zeta',S',T'} \bra{\begin{array}{c}
n_{14},n_{24},n_{34},n_{44}\cr
\zeta S, T 
\end{array}; \begin{array}{c}
n_{14}',n_{24}',n_{34}',n_{44}'\cr
\zeta' S' , T' 
\end{array}}\text{}\ket{\begin{array}{c}
n_{14}'',n_{24}'',n_{34}'',n_{44}''\cr
\zeta'' S'' , T'' 
\end{array}}_\eta \\
\times
\bra{\begin{array}{c}
n_{14},n_{24},n_{34},n_{44}\cr
\zeta S, T 
\end{array}; \begin{array}{c}
n_{14}',n_{24}',n_{34}',n_{44}'\cr
\zeta' S' , T' 
\end{array}}\text{}\ket{\begin{array}{c}
\tilde{n}_{14}'',\tilde{n}_{24}'',\tilde{n}_{34}'',\tilde{n}_{44}''\cr
\tilde{\zeta}'' S'' , T'' 
\end{array}}_{\tilde{\eta}} = \delta_{\zeta'',\tilde{\zeta}''}\delta_{\eta,\tilde{\eta}} \prod_{i=1}^4 \delta_{n_{i4}'',\tilde{n}_{i4}''},
    \label{eq:ortho1}
\end{multline}
and
\begin{multline}
    \sum_{n_{i4}'',\eta,\zeta''} \bra{\begin{array}{c}
n_{14},n_{24},n_{34},n_{44}\cr
\zeta S, T 
\end{array}; \begin{array}{c}
n_{14}',n_{24}',n_{34}',n_{44}'\cr
\zeta' S' , T' 
\end{array}}\text{}\ket{\begin{array}{c}
n_{14}'',n_{24}'',n_{34}'',n_{44}''\cr
\zeta'' S'' , T'' 
\end{array}}_\eta \\
\times
\bra{\begin{array}{c}
n_{14},n_{24},n_{34},n_{44}\cr
\tilde{\zeta} \tilde{S}, \tilde{T} 
\end{array}; \begin{array}{c}
n_{14}',n_{24}',n_{34}',n_{44}'\cr
\tilde{\zeta}' \tilde{S}', \tilde{T}'
\end{array}}\text{}\ket{\begin{array}{c}
n_{14}'',n_{24}'',n_{34}'',n_{44}''\cr
\zeta'' S'' , T'' 
\end{array}}_{\eta} = \delta_{\zeta,\tilde{\zeta}} \delta_{\zeta',\tilde{\zeta}'}  \delta_{S,\tilde{S}} \delta_{S',\tilde{S}'} \delta_{T,\tilde{T}} \delta_{T',\tilde{T}'}.
    \label{eq:ortho2}
\end{multline}
hold for the numerical values shown in the tables.

\begin{table}
    \centering
\caption{Wigner coefficients $\bra{\begin{array}{c}
8,4,2,0\cr
\zeta, S,T 
\end{array}; \begin{array}{c}
1,0,0,0\cr
\zeta'=1,S'=T'=1/2
\end{array}} \text{} \ket{\begin{array}{c}
9,4,2,0\cr
\zeta'',S''=T''=7/2
\end{array}}_{\eta=1}$}
\label{tab:eg1}
    \begin{tabular}{>{\centering\arraybackslash}p{1cm}>{\centering\arraybackslash}p{2cm}>{\centering\arraybackslash}p{2cm}>{\centering\arraybackslash}p{2cm}>{\centering\arraybackslash}p{2cm}>{\centering\arraybackslash}p{2cm}}
    \hline\hline
         $\zeta$&  $(S,T)$&  $\zeta''=1$&  $\zeta''=2$&  $\zeta''=3$& $\zeta''=4$\\
         \hline
         1
&  (4,4)&  -0.343019&  0.0612356&  -0.0179121& 0
\\ 
         1
&  (4,3)&  0.186572&  0.272502&  0& -0.137919
\\ 
         2
&  (4,3)&  0.100699&  -0.0179767&  -0.115251& 0.208514
\\ 
         1
&  (3,4)&  0.186572&  0.272502&  0& 0.137919
\\ 
         2
&  (3,4)&  -0.100699&  0.0179767&  0.115251& 0.208514
\\ 
         1
&  (3,3)&  0.886081&  0&  0& 0
\\ 
         2
&  (3,3)&  0.0163888&  0.0069387&  0.986407& 0
\\ 
         3
&  (3,3)&  0&  0&  0& 0.935414
\\
         4&  (3,3)&  -0.0838492&  0.920348&  -0.0107472& 0\\
           \hline\hline
    \end{tabular}

\end{table}

\begin{table}
    \centering
\caption{Wigner coefficients $\bra{\begin{array}{c}
8,4,2,0\cr
\zeta, S,T 
\end{array}; \begin{array}{c}
1,1,0,0\cr
\zeta'=1,S',T'
\end{array}} \text{} \ket{\begin{array}{c}
8,5,3,0\cr
\zeta'',S''=3,T''=2
\end{array}}_{\eta=1}$}
\label{tab:eg2}
    \begin{tabular}{>{\centering\arraybackslash}p{1cm}>{\centering\arraybackslash}p{1.6cm}>{\centering\arraybackslash}p{1.6cm}>{\centering\arraybackslash}p{2cm}>{\centering\arraybackslash}p{2cm}>{\centering\arraybackslash}p{2cm}>{\centering\arraybackslash}p{2cm}>{\centering\arraybackslash}p{2cm}}
    \hline\hline
         $\zeta$&  $(S,T)$&  $(S',T')$&  $\zeta''=1$&  $\zeta''=2$&  $\zeta''=3$&  $\zeta''=4$& $\zeta''=5$\\
         \hline
 1& (4,2)& (1,0)& 0& 0.0688505& -0.0620989& 0.194152&-0.118331
\\
 2& (4,2)& (1,0)& -0.486033& -0.000838901& 0.295462& 0.0743102&-0.0131602
\\
         3&  (4,2)&  (1,0)&  -0.0996271&  0.300124&  -0.255118&  0.0175132& 0.141745
\\
         1&  (3,2)&  (1,0)&  -0.0118857&  -0.155438&  -0.219715&  -0.289959& 0.0179052
\\
         2&  (3,2)&  (1,0)&  -0.0226953&  -0.362154&  0.268257&  0.0271416& -0.0764189
\\
         3&  (3,2)&  (1,0)&  0.045563&  -0.0733281&  -0.078353&  0.361672& -0.080103
\\
         4&  (3,2)&  (1,0)&  -0.0108016&  -0.0852314&  -0.0155587&  0.0999238& -0.541386
\\
         1&  (2,2)&  (1,0)&  0.173331&  0.126282&  -0.0495185&  0.047281& -0.0425521
\\
         2&  (2,2)&  (1,0)&  0.210996&  -0.0916816&  0.0179478&  -0.0220182& -0.12132
\\
         3&  (2,2)&  (1,0)&  0.154661&  -0.012515&  0.241789&  0.0735375& 0.256234
\\
 4& (2,2)& (1,0)& 0.00179472& 0.281128& -0.00862842& 0.171439&-0.224478
\\
 5& (2,2)& (1,0)& 0.0242476& 0.24691& 0.216476& -0.338965&-0.198874
\\
 1& (3,3)& (0,1)& 0.0282633& -0.0888694& -0.0344642& 0.25173&-0.035744
\\
 2& (3,3)& (0,1)& -0.637252& 0.019093& 0.282838& 0.0765627&-0.00393358
\\
 3& (3,3)& (0,1)& -0.196932& 0.147087& -0.458445& -0.0755306&0.0839061
\\
 4& (3,3)& (0,1)& -0.0718974& -0.380671& -0.12274& -0.134813&-0.160459
\\
 1& (3,2)& (0,1)& 0.291099& 0.235605& 0.335639& 0.365857&0.0237871
\\
 2& (3,2)& (0,1)& -0.145048& 0.409257& 0.0304183& -0.0289226&0.097562
\\
 3& (3,2)& (0,1)& 0.257869& 0.0227162& 0.311657& -0.444772&0.0391771
\\
 4& (3,2)& (0,1)& -0.00954735& -0.0222245& -0.0192093& 0.168838&0.421091
\\
 1& (3,1)& (0,1)& -0.117888& -0.0776855& -0.190022& -0.12693&-0.0258989
\\
 2& (3,1)& (0,1)& -0.121739& -0.0539785& 0.220995& -0.167361&-0.0341485
\\
 3& (3,1)& (0,1)& 0& -0.356882& -0.0533739& 0.278226&0.0567695
\\
 4& (3,1)& (0,1)& 0& 0.193875& -0.0606203& 0.0529075&-0.509759\\
 \hline\hline
    \end{tabular}

\end{table}

\subsection{Summary of the Algorithm for Computing U(4)$\supset$SU$_{\rm S}$(2)$\otimes$SU$_{\rm T}$(2)  Wigner Coefficients}

The procedure to compute all $\rm U_{ST}(4)$ Wigner coefficients associated with the tensor product of two U(4) irreps, $[n_{14},n_{24},n_{34},n_{44}] \otimes [n_{14}',n_{24}',n_{34}',n_{44}']$  can be summarized as follows,
\begin{enumerate}
    \item Find all coupled irreps $[n_{14}'',n_{24}'',n_{34}'',n_{44}'']$ that occur in the decomposition of the above tensor product,  as described in Subsection \ref{secIVA}.
    \item Calculate all $\rm U(4) \supset U(3)\supset U(2) \supset U(1)$ CGC's for the highest weight states (Subsection \ref{secIVC}) and then for all lower weight states (Subsection \ref{secIVD}).
    \item Carry out the procedure in Subsection \ref{secVIB} to find the expansion of the spin-isospin highest weight states, $\ket{[n_{14},n_{24},n_{34},n_{44}]\zeta S M_S=S, T M_T=T}$, $\ket{[n_{14}',n_{24}',n_{34}',n_{44}']\zeta' S' M_S'=S', T' M_T'=T'}$ and $\ket{[n_{14}'',n_{24}'',n_{34}'',n_{44}'']\zeta'' S'' M_S''=S'', T'' M_T''=T''}$, in their corresponding canonical Gelfand bases; and then the expansion of all lower weight states $\ket{[n_{14}',n_{24}',n_{34}',n_{44}']\zeta' S' M_S', T' M_T'}$  associated with the second irrep.
    \item Combine the results of the two previous steps to compute all Wigner coefficients as shown in Subsection \ref{secVIC}.
\end{enumerate}
Let us note that steps 2 and 3 are independent of each other, and therefore their order can be interchanged without affecting any results.

\section{Conclusion}
\label{secVII}

This paper introduces a relatively simple procedure for the evaluation of canonical $\rm U(4) \supset U(3)\supset U(2) \supset U(1)$ Clebsch-Gordan coefficients, which follows from a generic methodology for determining the U(N) CGC's that was advanced by Arne Alex et al. in 2011 \cite{Alex2011JMP}. Within this framework the  highest weight CGC's can be determined simultaneously from the solution of the null space of the three canonical raising generators of U(4), namely $E_{12}$, $E_{23}$ and $E_{34}$, while the lower weight CGC's can be obtained recursively through the action of the lowering generators $E_{21}$, $E_{32}$ and $E_{43}$. Since this methodology for computing CGC's in the canonical scheme is a purely mathematical exercise, the results are applicable to any physical system that follows all the properties of the U(4) algebra or respects U(4) symmetry (and canonical subgroup symmetries thereof). In addition, for completeness the calculation of Racah recoupling coefficients, which for reasons of clarity and simplicity are designated within this paper as 6-U(4) and 9-U(4) coefficients, is also presented.

\vskip.2cm

In addition, an application of the U(4) group to nuclear structure studies is addressed, which comes with a need to resolve the subgroup reduction $\rm U(4) \supset SU_S(2)\otimes SU_T(2)$, and thus requires the determination of any/all Wigner coupling coefficients associated thereto, which until now has not been satisfactorily resolved. However, this is no longer the case, since in this article new mathematical and computational methodologies, which have benefited from earlier efforts \cite{Pan2023EPJP,Pan2023NPA,Pan2024CPC}, are introduced; they sidestep prior sticky-wicket issues and provide for generic solutions to the U(4) nuclear physics challenge that is expected to lead to the development of a modern U(4) package that is similar in terms of functionalities to which was recently announced for U(3) \cite{Dang2024} and seems to meet all of the expectations associated with the advanced usage of Wigner's Supermultiplet Theory for carrying out next-generation nuclear structure studies. The heretofore missing ingredient central to this new approach is the availability of modern computational algorithms for resolving null space issues for both dense and sparse matrices. Furthermore, the latter can be fine-tuned to specific applications using modern programming languages. In fact, an early laptop version of a new C++ library for evaluating these coupling coefficients has already been developed and undergone elementary tests which show that any and all multiplicity issues are resolved correctly and that all dimensionality and orthogonality checks are satisfied. Most importantly, all of this bodes well in the support of future full implementations of Wigner's Supermultiplet Theory for carrying out advanced nuclear structure studies.

\vskip.2cm

In summary, the results of this paper show that advanced algebraic methodologies with support of modern computational algorithms (nullspace determinations), can be used to obtain a complete set of solutions for the coupling and recoupling coefficients for U(4). Additionally, the latter also seems to bode well regarding more advanced and perhaps even more universal applications of a similar type that may go well beyond its usage for just the U(4) unitary group, which is the main focus of this paper. To be more specific, in light of the emergence of state-of-the-art algebraic and computational technologies, it seems that the construction of a next-generation shell model theory, based upon the same principles as those that underpin the SA-NCSM, might actually be achievable within the current decade, and most certainly in time for a double jubilee (100 years) celebration of Wigher's 1937 proffering his supermultiplet symmetry picture, for which sufficient physical as well as financial resources will most certainly be required. Above all else, the authors of this paper wish to suggest that lying ahead is a very bright future for our gaining an even deeper appreciation for the key role that symmetries can and will most certainly play in our quest to uncover more about the true quantum nature of nuclear forces -- especially as this pertains to a far better untangling of the low-energy behavior of atomic nuclei that seems to be pervasive across the entire Chart of the Nuclides \footnote{The Chart of Nuclides: \href{https://www-nds.iaea.org/relnsd/vcharthtml/VChartHTML.html}{https://www-nds.iaea.org/relnsd/vcharthtml/VChartHTML.html}}.

%\bmhead{Acknowledgements}
\begin{acknowledgements}
{Support from the National Natural Science Foundation of China (12175097), from the Czech Science Foundation (22-14497S), from LSU through its Sponsored Research Rebate Program as well as under the LSU Foundation's Distinguished Research Professorship Program, and from the U.S. Department of Energy under grant number DE-SC0023532 are all acknowledged. This work also benefited from high performance computational resources provided by LSU (www.hpc.lsu.edu). Also, the lead author wishes to acknowledge financial assistance from the Student Ambassador Program of the American Physical Society, the organizers of the 2024 National Nuclear Physics Summer School at Indiana University Bloomington and the symposium in honor of the 75th Anniversary of the Shell Model to present early results of this work. In addition, the authors wish to acknowledge ongoing support from two LSU Professors, Kristina D. Launey and Alexis Mercenne, as well as Dr. Grigor H. Sargsyan from the Facility for Rare Isotope Beams at Michigan State University, and in particular for their feedback on this manuscript.}
\end{acknowledgements}

%\bmhead{Data availability}
\section*{Data availability}
Some simple Wigner $\rm U_{ST}(4)$ coupling coefficients, which were obtained using an early version of a C++ library for evaluating such coefficients, are given in Tables \ref{tab:eg1} and \ref{tab:eg2} in this article. The creation of a fully streamlined U(4) package remains under development - which is expected to be a far more robust feature that executes at higher processing speed. Nevertheless, the authors are prepared to entertain special requests for results of a reasonable set of U(4) coupling and recoupling coefficients, inclusive of the associated dimensionality checking, if so requested.

\bibliography{extracted_bib}

\begin{thebibliography}{55}
\expandafter\ifx\csname natexlab\endcsname\relax\def\natexlab#1{#1}\fi
\expandafter\ifx\csname bibnamefont\endcsname\relax
  \def\bibnamefont#1{#1}\fi
\expandafter\ifx\csname bibfnamefont\endcsname\relax
  \def\bibfnamefont#1{#1}\fi
\expandafter\ifx\csname citenamefont\endcsname\relax
  \def\citenamefont#1{#1}\fi
\expandafter\ifx\csname url\endcsname\relax
  \def\url#1{\texttt{#1}}\fi
\expandafter\ifx\csname urlprefix\endcsname\relax\def\urlprefix{URL }\fi
\providecommand{\bibinfo}[2]{#2}
\providecommand{\eprint}[2][]{\url{#2}}

\bibitem[{\citenamefont{Wigner}(1937)}]{Wigner1937PR}
\bibinfo{author}{\bibfnamefont{E.~P.} \bibnamefont{Wigner}}, \bibinfo{journal}{Physical Review} \textbf{\bibinfo{volume}{51}}, \bibinfo{pages}{106} (\bibinfo{year}{1937}).

\bibitem[{\citenamefont{Heisenberg}(1932)}]{Heisenberg1932ZP}
\bibinfo{author}{\bibfnamefont{W.}~\bibnamefont{Heisenberg}}, \bibinfo{journal}{Zeitschrift für Physik} \textbf{\bibinfo{volume}{77}}, \bibinfo{pages}{1} (\bibinfo{year}{1932}).

\bibitem[{\citenamefont{Cseh}(2014)}]{Cseh2014EPJ}
\bibinfo{author}{\bibfnamefont{J.}~\bibnamefont{Cseh}}, \bibinfo{journal}{EPJ Web of Conferences} \textbf{\bibinfo{volume}{78}}, \bibinfo{pages}{03002} (\bibinfo{year}{2014}).

\bibitem[{\citenamefont{Wigner}(1939)}]{Wigner1939PR}
\bibinfo{author}{\bibfnamefont{E.~P.} \bibnamefont{Wigner}}, \bibinfo{journal}{Physical Review} \textbf{\bibinfo{volume}{56}}, \bibinfo{pages}{519} (\bibinfo{year}{1939}).

\bibitem[{\citenamefont{Ikeda et~al.}(1962)\citenamefont{Ikeda, Fujii, and Fujita}}]{Ikeda1962PL}
\bibinfo{author}{\bibfnamefont{K.}~\bibnamefont{Ikeda}}, \bibinfo{author}{\bibfnamefont{S.}~\bibnamefont{Fujii}}, \bibnamefont{and} \bibinfo{author}{\bibfnamefont{F.~I.} \bibnamefont{Fujita}}, \bibinfo{journal}{Physics Letters} \textbf{\bibinfo{volume}{2}}, \bibinfo{pages}{169} (\bibinfo{year}{1962}).

\bibitem[{\citenamefont{Ikeda et~al.}(1963)\citenamefont{Ikeda, Fujii, and Fujita}}]{Ikeda1963PL}
\bibinfo{author}{\bibfnamefont{K.}~\bibnamefont{Ikeda}}, \bibinfo{author}{\bibfnamefont{S.}~\bibnamefont{Fujii}}, \bibnamefont{and} \bibinfo{author}{\bibfnamefont{F.~I.} \bibnamefont{Fujita}}, \bibinfo{journal}{Physics Letters} \textbf{\bibinfo{volume}{3}}, \bibinfo{pages}{271} (\bibinfo{year}{1963}).

\bibitem[{\citenamefont{Fujita et~al.}(1964)\citenamefont{Fujita, Fujii, and Ikeda}}]{Fujita1964PR}
\bibinfo{author}{\bibfnamefont{J.-I.} \bibnamefont{Fujita}}, \bibinfo{author}{\bibfnamefont{S.}~\bibnamefont{Fujii}}, \bibnamefont{and} \bibinfo{author}{\bibfnamefont{K.}~\bibnamefont{Ikeda}}, \bibinfo{journal}{Physical Review} \textbf{\bibinfo{volume}{133}}, \bibinfo{pages}{B549} (\bibinfo{year}{1964}).

\bibitem[{\citenamefont{Fujita and Ikeda}(1965)}]{Fujita1965NP}
\bibinfo{author}{\bibfnamefont{J.-I.} \bibnamefont{Fujita}} \bibnamefont{and} \bibinfo{author}{\bibfnamefont{K.}~\bibnamefont{Ikeda}}, \bibinfo{journal}{Nuclear Physics} \textbf{\bibinfo{volume}{67}}, \bibinfo{pages}{145} (\bibinfo{year}{1965}).

\bibitem[{\citenamefont{Fujita and et~al}(2019)}]{Fujita2019PRC}
\bibinfo{author}{\bibfnamefont{H.}~\bibnamefont{Fujita}} \bibnamefont{and} \bibinfo{author}{\bibnamefont{et~al}}, \bibinfo{journal}{Physical Review C} \textbf{\bibinfo{volume}{100}}, \bibinfo{pages}{034618} (\bibinfo{year}{2019}).

\bibitem[{\citenamefont{Gaponov and Lutostansky}(2010)}]{Gaponov2010PoAN}
\bibinfo{author}{\bibfnamefont{Y.~V.} \bibnamefont{Gaponov}} \bibnamefont{and} \bibinfo{author}{\bibfnamefont{Y.~S.} \bibnamefont{Lutostansky}}, \bibinfo{journal}{Physics of Atomic Nuclei} \textbf{\bibinfo{volume}{73}}, \bibinfo{pages}{1360} (\bibinfo{year}{2010}).

\bibitem[{\citenamefont{Lutostansky and Tikhonov}(2016)}]{Lutostansky2016EPJ}
\bibinfo{author}{\bibfnamefont{Y.~S.} \bibnamefont{Lutostansky}} \bibnamefont{and} \bibinfo{author}{\bibfnamefont{V.~N.} \bibnamefont{Tikhonov}}, \bibinfo{journal}{EPJ Web of Conferences} \textbf{\bibinfo{volume}{107}}, \bibinfo{pages}{06004} (\bibinfo{year}{2016}).

\bibitem[{\citenamefont{Kaplan and Savage}(1996)}]{Kaplan1996PLB}
\bibinfo{author}{\bibfnamefont{D.~B.} \bibnamefont{Kaplan}} \bibnamefont{and} \bibinfo{author}{\bibfnamefont{M.~J.} \bibnamefont{Savage}}, \bibinfo{journal}{Physics Letters B} \textbf{\bibinfo{volume}{365}}, \bibinfo{pages}{244} (\bibinfo{year}{1996}).

\bibitem[{\citenamefont{Kaplan and Manohar}(1997)}]{Kaplan1997PRC}
\bibinfo{author}{\bibfnamefont{D.~B.} \bibnamefont{Kaplan}} \bibnamefont{and} \bibinfo{author}{\bibfnamefont{A.~V.} \bibnamefont{Manohar}}, \bibinfo{journal}{Physical Review C} \textbf{\bibinfo{volume}{56}}, \bibinfo{pages}{56} (\bibinfo{year}{1997}).

\bibitem[{\citenamefont{Cord\'on and Arriola}(2008)}]{Cordon2008PRC}
\bibinfo{author}{\bibfnamefont{A.~C.} \bibnamefont{Cord\'on}} \bibnamefont{and} \bibinfo{author}{\bibfnamefont{E.~R.} \bibnamefont{Arriola}}, \bibinfo{journal}{Physical Review C} \textbf{\bibinfo{volume}{78}}, \bibinfo{pages}{054002} (\bibinfo{year}{2008}).

\bibitem[{\citenamefont{Mehen et~al.}(1999)\citenamefont{Mehen, Stewart, and Wise}}]{Mehen1999PRL}
\bibinfo{author}{\bibfnamefont{T.}~\bibnamefont{Mehen}}, \bibinfo{author}{\bibfnamefont{I.~W.} \bibnamefont{Stewart}}, \bibnamefont{and} \bibinfo{author}{\bibfnamefont{M.~B.} \bibnamefont{Wise}}, \bibinfo{journal}{Physical Review Letters} \textbf{\bibinfo{volume}{83}}, \bibinfo{pages}{931} (\bibinfo{year}{1999}).

\bibitem[{\citenamefont{Beane and et~al}(2013)}]{Beane2013PRC}
\bibinfo{author}{\bibfnamefont{S.~R.} \bibnamefont{Beane}} \bibnamefont{and} \bibinfo{author}{\bibnamefont{et~al}}, \bibinfo{journal}{Physical Review C} \textbf{\bibinfo{volume}{88}}, \bibinfo{pages}{024003} (\bibinfo{year}{2013}).

\bibitem[{\citenamefont{Lu and et~al}(2019)}]{Lu2019PLB}
\bibinfo{author}{\bibfnamefont{B.-N.} \bibnamefont{Lu}} \bibnamefont{and} \bibinfo{author}{\bibnamefont{et~al}}, \bibinfo{journal}{Physics Letters B} \textbf{\bibinfo{volume}{797}}, \bibinfo{pages}{134863} (\bibinfo{year}{2019}).

\bibitem[{\citenamefont{Beane et~al.}(2019)\citenamefont{Beane, Kaplan, Klco, and Savage}}]{Beane2019PRL}
\bibinfo{author}{\bibfnamefont{S.~R.} \bibnamefont{Beane}}, \bibinfo{author}{\bibfnamefont{D.~B.} \bibnamefont{Kaplan}}, \bibinfo{author}{\bibfnamefont{N.}~\bibnamefont{Klco}}, \bibnamefont{and} \bibinfo{author}{\bibfnamefont{M.~J.} \bibnamefont{Savage}}, \bibinfo{journal}{Physical Review Letters} \textbf{\bibinfo{volume}{122}}, \bibinfo{pages}{102001} (\bibinfo{year}{2019}).

\bibitem[{\citenamefont{Liu et~al.}(2023)\citenamefont{Liu, Low, and Mehen}}]{Liu2023PRC}
\bibinfo{author}{\bibfnamefont{Q.}~\bibnamefont{Liu}}, \bibinfo{author}{\bibfnamefont{I.}~\bibnamefont{Low}}, \bibnamefont{and} \bibinfo{author}{\bibfnamefont{T.}~\bibnamefont{Mehen}}, \bibinfo{journal}{Physical Review C} \textbf{\bibinfo{volume}{107}}, \bibinfo{pages}{025204} (\bibinfo{year}{2023}).

\bibitem[{\citenamefont{Miller}(2023)}]{Miller2023PRC}
\bibinfo{author}{\bibfnamefont{G.~A.} \bibnamefont{Miller}}, \bibinfo{journal}{Physical Review C} \textbf{\bibinfo{volume}{108}}, \bibinfo{pages}{L031002} (\bibinfo{year}{2023}).

\bibitem[{\citenamefont{Kota and Sahu}(2024)}]{Kota2024PScr}
\bibinfo{author}{\bibfnamefont{V.~K.~B.} \bibnamefont{Kota}} \bibnamefont{and} \bibinfo{author}{\bibfnamefont{R.}~\bibnamefont{Sahu}}, \bibinfo{journal}{Physica Scripta} \textbf{\bibinfo{volume}{99}}, \bibinfo{pages}{065306} (\bibinfo{year}{2024}).

\bibitem[{\citenamefont{Elliott}(1958{\natexlab{a}})}]{Elliott1958PRSA}
\bibinfo{author}{\bibfnamefont{J.}~\bibnamefont{Elliott}}, \bibinfo{journal}{Proceedings of the Royal Society A} \textbf{\bibinfo{volume}{245}}, \bibinfo{pages}{128} (\bibinfo{year}{1958}{\natexlab{a}}).

\bibitem[{\citenamefont{Elliott}(1958{\natexlab{b}})}]{Elliott1958PRSA-II}
\bibinfo{author}{\bibfnamefont{J.~P.} \bibnamefont{Elliott}}, \bibinfo{journal}{Proceedings of the Royal Society A} \textbf{\bibinfo{volume}{245}}, \bibinfo{pages}{562} (\bibinfo{year}{1958}{\natexlab{b}}).

\bibitem[{\citenamefont{Draayer et~al.}(1989)\citenamefont{Draayer, Park, and Castaños}}]{Draayer1989PRL}
\bibinfo{author}{\bibfnamefont{J.~P.} \bibnamefont{Draayer}}, \bibinfo{author}{\bibfnamefont{S.~C.} \bibnamefont{Park}}, \bibnamefont{and} \bibinfo{author}{\bibfnamefont{O.}~\bibnamefont{Castaños}}, \bibinfo{journal}{Physical Review Letters} \textbf{\bibinfo{volume}{62}}, \bibinfo{pages}{20} (\bibinfo{year}{1989}).

\bibitem[{\citenamefont{Rosensteel and Rowe}(1977)}]{Rosensteel1977AnnP}
\bibinfo{author}{\bibfnamefont{G.}~\bibnamefont{Rosensteel}} \bibnamefont{and} \bibinfo{author}{\bibfnamefont{D.}~\bibnamefont{Rowe}}, \bibinfo{journal}{Annals of Physics} \textbf{\bibinfo{volume}{104}}, \bibinfo{pages}{134} (\bibinfo{year}{1977}).

\bibitem[{\citenamefont{Rosensteel and Rowe}(1980)}]{Rosensteel1980AnnP}
\bibinfo{author}{\bibfnamefont{G.}~\bibnamefont{Rosensteel}} \bibnamefont{and} \bibinfo{author}{\bibfnamefont{D.}~\bibnamefont{Rowe}}, \bibinfo{journal}{Annals of Physics} \textbf{\bibinfo{volume}{126}}, \bibinfo{pages}{343} (\bibinfo{year}{1980}).

\bibitem[{\citenamefont{Dytrych et~al.}(2007)\citenamefont{Dytrych, Sviratcheva, Bahri, Draayer, and Vary}}]{Dytrych2007PRL}
\bibinfo{author}{\bibfnamefont{T.}~\bibnamefont{Dytrych}}, \bibinfo{author}{\bibfnamefont{K.}~\bibnamefont{Sviratcheva}}, \bibinfo{author}{\bibfnamefont{C.}~\bibnamefont{Bahri}}, \bibinfo{author}{\bibfnamefont{J.}~\bibnamefont{Draayer}}, \bibnamefont{and} \bibinfo{author}{\bibfnamefont{J.}~\bibnamefont{Vary}}, \bibinfo{journal}{Physical Review Letters} \textbf{\bibinfo{volume}{98}}, \bibinfo{pages}{162503} (\bibinfo{year}{2007}).

\bibitem[{\citenamefont{Dytrych and et~al}(2020)}]{Dytrych2020PRL}
\bibinfo{author}{\bibfnamefont{T.}~\bibnamefont{Dytrych}} \bibnamefont{and} \bibinfo{author}{\bibnamefont{et~al}}, \bibinfo{journal}{Physical Review Letters} \textbf{\bibinfo{volume}{124}}, \bibinfo{pages}{042501} (\bibinfo{year}{2020}).

\bibitem[{\citenamefont{Cseh}(2021)}]{Cseh2021PRC}
\bibinfo{author}{\bibfnamefont{J.}~\bibnamefont{Cseh}}, \bibinfo{journal}{Physical Review C} \textbf{\bibinfo{volume}{103}}, \bibinfo{pages}{064322} (\bibinfo{year}{2021}).

\bibitem[{\citenamefont{Raju et~al.}(1973)\citenamefont{Raju, Draayer, and Hecht}}]{Raju1973NPA}
\bibinfo{author}{\bibfnamefont{R.~D.~R.} \bibnamefont{Raju}}, \bibinfo{author}{\bibfnamefont{J.~P.} \bibnamefont{Draayer}}, \bibnamefont{and} \bibinfo{author}{\bibfnamefont{K.~T.} \bibnamefont{Hecht}}, \bibinfo{journal}{Nuclear Physics A} \textbf{\bibinfo{volume}{202}}, \bibinfo{pages}{433} (\bibinfo{year}{1973}).

\bibitem[{\citenamefont{Bonatsos et~al.}(2017)\citenamefont{Bonatsos, Assimakis, Minkov, Martinou, Cakirli, Casten, and Blaum}}]{Bonatsos2017PRC}
\bibinfo{author}{\bibfnamefont{D.}~\bibnamefont{Bonatsos}}, \bibinfo{author}{\bibfnamefont{I.~E.} \bibnamefont{Assimakis}}, \bibinfo{author}{\bibfnamefont{N.}~\bibnamefont{Minkov}}, \bibinfo{author}{\bibfnamefont{A.}~\bibnamefont{Martinou}}, \bibinfo{author}{\bibfnamefont{R.~B.} \bibnamefont{Cakirli}}, \bibinfo{author}{\bibfnamefont{R.~F.} \bibnamefont{Casten}}, \bibnamefont{and} \bibinfo{author}{\bibfnamefont{K.}~\bibnamefont{Blaum}}, \bibinfo{journal}{Physical Review C} \textbf{\bibinfo{volume}{95}}, \bibinfo{pages}{064325} (\bibinfo{year}{2017}).

\bibitem[{\citenamefont{Cseh}(2020)}]{Cseh2020PRC}
\bibinfo{author}{\bibfnamefont{J.}~\bibnamefont{Cseh}}, \bibinfo{journal}{Physical Review C} \textbf{\bibinfo{volume}{101}}, \bibinfo{pages}{054306} (\bibinfo{year}{2020}).

\bibitem[{\citenamefont{Pan et~al.}(2023{\natexlab{a}})\citenamefont{Pan, Wu, Li, Zhang, Dai, and Draayer}}]{Pan2023EPJP}
\bibinfo{author}{\bibfnamefont{F.}~\bibnamefont{Pan}}, \bibinfo{author}{\bibfnamefont{Y.}~\bibnamefont{Wu}}, \bibinfo{author}{\bibfnamefont{A.}~\bibnamefont{Li}}, \bibinfo{author}{\bibfnamefont{Y.}~\bibnamefont{Zhang}}, \bibinfo{author}{\bibfnamefont{L.}~\bibnamefont{Dai}}, \bibnamefont{and} \bibinfo{author}{\bibfnamefont{J.~P.} \bibnamefont{Draayer}}, \bibinfo{journal}{The European Physical Journal Plus} \textbf{\bibinfo{volume}{138}}, \bibinfo{pages}{662} (\bibinfo{year}{2023}{\natexlab{a}}).

\bibitem[{\citenamefont{Pan et~al.}(2023{\natexlab{b}})\citenamefont{Pan, Dai, and Draayer}}]{Pan2023NPA}
\bibinfo{author}{\bibfnamefont{F.}~\bibnamefont{Pan}}, \bibinfo{author}{\bibfnamefont{L.}~\bibnamefont{Dai}}, \bibnamefont{and} \bibinfo{author}{\bibfnamefont{J.~P.} \bibnamefont{Draayer}}, \bibinfo{journal}{Nuclear Physics A} \textbf{\bibinfo{volume}{1040}}, \bibinfo{pages}{122746} (\bibinfo{year}{2023}{\natexlab{b}}).

\bibitem[{\citenamefont{{F. Pan and L. Dai and J. P. Draayer}}(2024)}]{Pan2024CPC}
\bibinfo{author}{\bibnamefont{{F. Pan and L. Dai and J. P. Draayer}}}, \bibinfo{journal}{Computer Physics Communications} \textbf{\bibinfo{volume}{296}}, \bibinfo{pages}{109025} (\bibinfo{year}{2024}).

\bibitem[{\citenamefont{Alex et~al.}(2011)\citenamefont{Alex, Kalus, Huckleberry, and von Delft}}]{Alex2011JMP}
\bibinfo{author}{\bibfnamefont{A.}~\bibnamefont{Alex}}, \bibinfo{author}{\bibfnamefont{M.}~\bibnamefont{Kalus}}, \bibinfo{author}{\bibfnamefont{A.}~\bibnamefont{Huckleberry}}, \bibnamefont{and} \bibinfo{author}{\bibfnamefont{J.}~\bibnamefont{von Delft}}, \bibinfo{journal}{Journal of Mathematical Physics} \textbf{\bibinfo{volume}{52}}, \bibinfo{pages}{023507} (\bibinfo{year}{2011}).

\bibitem[{\citenamefont{Akiyama and Draayer}(1973)}]{Akiyama1973CPC}
\bibinfo{author}{\bibfnamefont{Y.}~\bibnamefont{Akiyama}} \bibnamefont{and} \bibinfo{author}{\bibfnamefont{J.~P.} \bibnamefont{Draayer}}, \bibinfo{journal}{Computer Physics Communications} \textbf{\bibinfo{volume}{5}}, \bibinfo{pages}{405} (\bibinfo{year}{1973}).

\bibitem[{\citenamefont{Draayer and Akiyama}(1973)}]{Draayer1973JMP}
\bibinfo{author}{\bibfnamefont{J.~P.} \bibnamefont{Draayer}} \bibnamefont{and} \bibinfo{author}{\bibfnamefont{Y.}~\bibnamefont{Akiyama}}, \bibinfo{journal}{Journal of Mathematical Physics} \textbf{\bibinfo{volume}{14}}, \bibinfo{pages}{1904} (\bibinfo{year}{1973}).

\bibitem[{\citenamefont{Dytrych et~al.}(2021)\citenamefont{Dytrych, Langr, Draayer, Launey, and Gazda}}]{Dytrych2021CPC}
\bibinfo{author}{\bibfnamefont{T.}~\bibnamefont{Dytrych}}, \bibinfo{author}{\bibfnamefont{D.}~\bibnamefont{Langr}}, \bibinfo{author}{\bibfnamefont{J.~P.} \bibnamefont{Draayer}}, \bibinfo{author}{\bibfnamefont{K.~D.} \bibnamefont{Launey}}, \bibnamefont{and} \bibinfo{author}{\bibfnamefont{D.}~\bibnamefont{Gazda}}, \bibinfo{journal}{Computer Physics Communications} \textbf{\bibinfo{volume}{269}}, \bibinfo{pages}{108137} (\bibinfo{year}{2021}).

\bibitem[{\citenamefont{Dang et~al.}(2024)\citenamefont{Dang, Draayer, Pan, and Becker}}]{Dang2024}
\bibinfo{author}{\bibfnamefont{P.}~\bibnamefont{Dang}}, \bibinfo{author}{\bibfnamefont{J.~P.} \bibnamefont{Draayer}}, \bibinfo{author}{\bibfnamefont{F.}~\bibnamefont{Pan}}, \bibnamefont{and} \bibinfo{author}{\bibfnamefont{K.~S.} \bibnamefont{Becker}}, \emph{\bibinfo{title}{{New Procedure for Evaluation of $\rm U(3)$ Coupling and Recoupling Coefficients}}} (\bibinfo{year}{2024}), \eprint{2405.06843}.

\bibitem[{\citenamefont{Pan et~al.}(2016)\citenamefont{Pan, Yuan, Launey, and Draayer}}]{Pan2016NPA}
\bibinfo{author}{\bibfnamefont{F.}~\bibnamefont{Pan}}, \bibinfo{author}{\bibfnamefont{S.}~\bibnamefont{Yuan}}, \bibinfo{author}{\bibfnamefont{K.~D.} \bibnamefont{Launey}}, \bibnamefont{and} \bibinfo{author}{\bibfnamefont{J.~P.} \bibnamefont{Draayer}}, \bibinfo{journal}{Nuclear Physics A} \textbf{\bibinfo{volume}{952}}, \bibinfo{pages}{70} (\bibinfo{year}{2016}).

\bibitem[{\citenamefont{Hecht and Pang}(1969)}]{Hecht1969JMP}
\bibinfo{author}{\bibfnamefont{K.~T.} \bibnamefont{Hecht}} \bibnamefont{and} \bibinfo{author}{\bibfnamefont{S.~P.} \bibnamefont{Pang}}, \bibinfo{journal}{Journal of Mathematical Physics} \textbf{\bibinfo{volume}{10}}, \bibinfo{pages}{1571} (\bibinfo{year}{1969}).

\bibitem[{\citenamefont{Draayer}(1970)}]{Draayer1970JMP}
\bibinfo{author}{\bibfnamefont{J.~P.} \bibnamefont{Draayer}}, \bibinfo{journal}{Journal of Mathematical Physics} \textbf{\bibinfo{volume}{11}}, \bibinfo{pages}{3225} (\bibinfo{year}{1970}).

\bibitem[{\citenamefont{Partensky and Maguin}(1987)}]{Partensky1978JMP}
\bibinfo{author}{\bibfnamefont{A.}~\bibnamefont{Partensky}} \bibnamefont{and} \bibinfo{author}{\bibfnamefont{C.}~\bibnamefont{Maguin}}, \bibinfo{journal}{Journal of Mathematical Physics} \textbf{\bibinfo{volume}{19}}, \bibinfo{pages}{511} (\bibinfo{year}{1987}).

\bibitem[{\citenamefont{Rowe and Repka}(1997)}]{Rowe1997FoP}
\bibinfo{author}{\bibfnamefont{D.~J.} \bibnamefont{Rowe}} \bibnamefont{and} \bibinfo{author}{\bibfnamefont{J.}~\bibnamefont{Repka}}, \bibinfo{journal}{Foundations of Physics} \textbf{\bibinfo{volume}{27}}, \bibinfo{pages}{1179} (\bibinfo{year}{1997}).

\bibitem[{\citenamefont{Louck and Biedenharn}(1970)}]{Louck1970JMP}
\bibinfo{author}{\bibfnamefont{J.~D.} \bibnamefont{Louck}} \bibnamefont{and} \bibinfo{author}{\bibfnamefont{L.~C.} \bibnamefont{Biedenharn}}, \bibinfo{journal}{Journal of Mathematical Physics} \textbf{\bibinfo{volume}{11}}, \bibinfo{pages}{2368} (\bibinfo{year}{1970}).

\bibitem[{\citenamefont{Biedenharn et~al.}(1972)\citenamefont{Biedenharn, Louck, Chacón, and Ciftan}}]{Biedenharn1972JMP1}
\bibinfo{author}{\bibfnamefont{L.~C.} \bibnamefont{Biedenharn}}, \bibinfo{author}{\bibfnamefont{J.~D.} \bibnamefont{Louck}}, \bibinfo{author}{\bibfnamefont{E.}~\bibnamefont{Chacón}}, \bibnamefont{and} \bibinfo{author}{\bibfnamefont{M.}~\bibnamefont{Ciftan}}, \bibinfo{journal}{Journal of Mathematical Physics} \textbf{\bibinfo{volume}{13}}, \bibinfo{pages}{1957} (\bibinfo{year}{1972}).

\bibitem[{\citenamefont{Biedenharn and Louck}(1972)}]{Biedenharn1972JMP2}
\bibinfo{author}{\bibfnamefont{L.~C.} \bibnamefont{Biedenharn}} \bibnamefont{and} \bibinfo{author}{\bibfnamefont{J.~D.} \bibnamefont{Louck}}, \bibinfo{journal}{Journal of Mathematical Physics} \textbf{\bibinfo{volume}{13}}, \bibinfo{pages}{1985} (\bibinfo{year}{1972}).

\bibitem[{\citenamefont{Louck and Biedenharn}(1973)}]{Louck1973JMP}
\bibinfo{author}{\bibfnamefont{J.~D.} \bibnamefont{Louck}} \bibnamefont{and} \bibinfo{author}{\bibfnamefont{L.~C.} \bibnamefont{Biedenharn}}, \bibinfo{journal}{Journal of Mathematical Physics} \textbf{\bibinfo{volume}{14}}, \bibinfo{pages}{1336} (\bibinfo{year}{1973}).

\bibitem[{\citenamefont{Kuhn and Walliser}(2008)}]{Kuhn2008CPC}
\bibinfo{author}{\bibfnamefont{M.}~\bibnamefont{Kuhn}} \bibnamefont{and} \bibinfo{author}{\bibfnamefont{H.}~\bibnamefont{Walliser}}, \bibinfo{journal}{Computer Physics Communications} \textbf{\bibinfo{volume}{179}}, \bibinfo{pages}{733} (\bibinfo{year}{2008}).

\bibitem[{\citenamefont{Gel'fand and Zetlin}(1950)}]{Gelfand1950}
\bibinfo{author}{\bibfnamefont{L.~M.} \bibnamefont{Gel'fand}} \bibnamefont{and} \bibinfo{author}{\bibfnamefont{M.~L.} \bibnamefont{Zetlin}}, \bibinfo{journal}{Dokl. Akad. Nauk. SSSR} \textbf{\bibinfo{volume}{71}}, \bibinfo{pages}{825} (\bibinfo{year}{1950}).

\bibitem[{\citenamefont{Cseh}(1992)}]{Cseh1992PLB}
\bibinfo{author}{\bibfnamefont{J.}~\bibnamefont{Cseh}}, \bibinfo{journal}{Physics Letters B} \textbf{\bibinfo{volume}{281}}, \bibinfo{pages}{173} (\bibinfo{year}{1992}).

\bibitem[{\citenamefont{Cseh and L\'evai}(1994)}]{Cseh1994AnnP}
\bibinfo{author}{\bibfnamefont{J.}~\bibnamefont{Cseh}} \bibnamefont{and} \bibinfo{author}{\bibfnamefont{G.}~\bibnamefont{L\'evai}}, \bibinfo{journal}{Annals of Physics} \textbf{\bibinfo{volume}{230}}, \bibinfo{pages}{165} (\bibinfo{year}{1994}).

\bibitem[{\citenamefont{Dang et~al.}(2023)\citenamefont{Dang, Riczu, and Cseh}}]{Dang2023PRC}
\bibinfo{author}{\bibfnamefont{P.}~\bibnamefont{Dang}}, \bibinfo{author}{\bibfnamefont{G.}~\bibnamefont{Riczu}}, \bibnamefont{and} \bibinfo{author}{\bibfnamefont{J.}~\bibnamefont{Cseh}}, \bibinfo{journal}{Physical Review C} \textbf{\bibinfo{volume}{107}}, \bibinfo{pages}{044315} (\bibinfo{year}{2023}).

\bibitem[{\citenamefont{Racah}(1949)}]{Racah1949RMP}
\bibinfo{author}{\bibfnamefont{G.}~\bibnamefont{Racah}}, \bibinfo{journal}{Reviews of Modern Physics} \textbf{\bibinfo{volume}{21}}, \bibinfo{pages}{3} (\bibinfo{year}{1949}).

\end{thebibliography}

\appendix

\section{Action of the U(4) raising generators}
\label{appendA}
\begin{align}
    e_{12}(G) &= \sqrt{(n_{12}-n_{11})(n_{11}-n_{22}+1)}, \nonumber 
\end{align}
\begin{align}
    e_{23,12}(G) &= \sqrt{\frac{(n_{13}-n_{12})(n_{12}-n_{23}+1)(n_{12}-n_{33}+2)(n_{12}-n_{11}+1)}{(n_{12}-n_{22}+2)(n_{12}-n_{22}+1)}} \nonumber 
\end{align}
\begin{align}
    e_{23,22}(G) &= \sqrt{\frac{(n_{13}-n_{22}+1)(n_{23}-n_{22})(n_{22}-n_{33}+1)(n_{11}-n_{22})}{(n_{12}-n_{22}+1)(n_{12}-n_{22})}},  \nonumber 
\end{align}
\begin{align}
    e_{13,12}(G) &= \sqrt{\frac{(n_{13}-n_{12})(n_{12}-n_{23}+1)(n_{12}-n_{33}+2)(n_{11}-n_{22}+1)}{(n_{12}-n_{22}+2)(n_{12}-n_{22}+1)}}, \nonumber
\end{align}
\begin{align}
    e_{13,22} (G) &= - \sqrt{\frac{(n_{13}-n_{22}+1)(n_{23}-n_{22})(n_{22}-n_{33}+1)(n_{12}-n_{11})}{(n_{12}-n_{22}+1)(n_{12}-n_{22})}}, \nonumber 
\end{align}
\begin{align}
    e_{34,13}(G) &= \sqrt{\frac{(n_{14}-n_{13})(n_{13}-n_{24}+1)(n_{13}-n_{34}+2)(n_{13}-n_{44}+3)(n_{13}-n_{12}+1)(n_{13}-n_{22}+2)}{(n_{13}-n_{23}+2)(n_{13}-n_{23}+1)(n_{13}-n_{33}+3)(n_{13}-n_{33}+2)}} \nonumber 
\end{align}
\begin{align}
    e_{34,23}(G) &= \sqrt{\frac{(n_{14}-n_{23}+1)(n_{24}-n_{23})(n_{23}-n_{34}+1)(n_{23}-n_{44}+2)(n_{12}-n_{23})(n_{23}-n_{22}+1)}{(n_{13}-n_{23}+1)(n_{13}-n_{23})(n_{23}-n_{33}+2)(n_{23}-n_{33}+1)}} \nonumber 
\end{align}
\begin{align}
    e_{34,33}(G) &= \sqrt{\frac{(n_{14}-n_{33}+2)(n_{24}-n_{33}+1)(n_{34}-n_{33})(n_{33}-n_{44}+1)(n_{12}-n_{33}+1)(n_{22}-n_{33})}{(n_{13}-n_{33}+2)(n_{13}-n_{33}+1)(n_{23}-n_{33}+1)(n_{23}-n_{33})}} \nonumber 
\end{align}
\begin{align}
    e_{24,13,12}(G) &= \sqrt{\frac{(n_{14}-n_{13})(n_{13}-n_{24}+1)(n_{13}-n_{34}+2)(n_{13}-n_{44}+3)(n_{12}-n_{23}+1)}{(n_{13}-n_{23}+2)(n_{13}-n_{23}+1)(n_{13}-n_{33}+3)(n_{13}-n_{33}+2)}} 
    \nonumber \\
    & \times \sqrt{\frac{(n_{12}-n_{33}+2)(n_{13}-n_{22}+2)(n_{12}-n_{11}+1)}{(n_{12}-n_{22}+2)(n_{12}-n_{22}+1)}} \nonumber 
\end{align}
\begin{align}
    e_{24,13,22}(G) &= \sqrt{\frac{(n_{14}-n_{13})(n_{13}-n_{24}+1)(n_{13}-n_{34}+2)(n_{13}-n_{44}+3)(n_{23}-n_{22})}{(n_{13}-n_{23}+2)(n_{13}-n_{23}+1)(n_{13}-n_{33}+3)(n_{13}-n_{33}+2)}} \nonumber \\
    & \times \sqrt{\frac{(n_{22}-n_{33}+1)(n_{13}-n_{12}+1)(n_{11}-n_{22})}{(n_{12}-n_{22}+1)(n_{12}-n_{22})}} \nonumber 
\end{align}
\begin{align}
    e_{24,23,12}(G) &= -\sqrt{\frac{(n_{14}-n_{23}+1)(n_{24}-n_{23})(n_{23}-n_{34}+1)(n_{23}-n_{44}+2)(n_{13}-n_{12})}{(n_{13}-n_{23}+1)(n_{13}-n_{23})(n_{23}-n_{33}+2)(n_{23}-n_{33}+1)}} \nonumber \\
    & \times \sqrt{\frac{(n_{12}-n_{33}+2)(n_{23}-n_{22}+1)(n_{12}-n_{11}+1)}{(n_{12}-n_{22}+2)(n_{12}-n_{22}+1)}} \nonumber 
\end{align}
\begin{align}
    e_{24,23,22}(G) &= \sqrt{\frac{(n_{14}-n_{23}+1)(n_{24}-n_{23})(n_{23}-n_{34}+1)(n_{23}-n_{44}+2)(n_{13}-n_{22}+1)}{(n_{13}-n_{23}+1)(n_{13}-n_{23})(n_{23}-n_{33}+2)(n_{23}-n_{33}+1)}} \nonumber \\
    & \times \sqrt{\frac{(n_{22}-n_{33}+1)(n_{12}-n_{23})(n_{11}-n_{22})}{(n_{12}-n_{22}+1)(n_{12}-n_{22})}} \nonumber 
\end{align}
\begin{align}
    e_{24,33,12}(G) &= -\sqrt{\frac{(n_{14}-n_{33}+2)(n_{24}-n_{33}+1)(n_{34}-n_{33})(n_{33}-n_{44}+1)(n_{13}-n_{12})}{(n_{13}-n_{33}+2)(n_{13}-n_{33}+1)(n_{23}-n_{33}+1)(n_{23}-n_{33})}} \nonumber \\
    & \times \sqrt{\frac{(n_{12}-n_{23}+1)(n_{22}-n_{33})(n_{12}-n_{11}+1)}{(n_{12}-n_{22}+2)(n_{12}-n_{22}+1)}} \nonumber 
\end{align}
\begin{align}
    e_{24,33,22}(G) &= -\sqrt{\frac{(n_{14}-n_{33}+2)(n_{24}-n_{33}+1)(n_{34}-n_{33})(n_{33}-n_{44}+1)(n_{13}-n_{22}+1)}{(n_{13}-n_{33}+2)(n_{13}-n_{33}+1)(n_{23}-n_{33}+1)(n_{23}-n_{33})}} \nonumber \\
    & \times \sqrt{\frac{(n_{23}-n_{22})(n_{12}-n_{33}+1)(n_{11}-n_{22})}{(n_{12}-n_{22}+1)(n_{12}-n_{22})}} \nonumber
\end{align}
\begin{align}
    e_{14,13,12}(G) &= \sqrt{\frac{(n_{14}-n_{13})(n_{13}-n_{24}+1)(n_{13}-n_{34}+2)(n_{13}-n_{44}+3)(n_{12}-n_{23}+1)}{(n_{13}-n_{23}+2)(n_{13}-n_{23}+1)(n_{13}-n_{33}+3)(n_{13}-n_{33}+2)}} 
    \nonumber \\
    & \times \sqrt{\frac{(n_{12}-n_{33}+2)(n_{13}-n_{22}+2)(n_{11}-n_{22}+1)}{(n_{12}-n_{22}+2)(n_{12}-n_{22}+1)}} \nonumber 
\end{align}
\begin{align}
    e_{14,13,22}(G) &= -\sqrt{\frac{(n_{14}-n_{13})(n_{13}-n_{24}+1)(n_{13}-n_{34}+2)(n_{13}-n_{44}+3)(n_{23}-n_{22})}{(n_{13}-n_{23}+2)(n_{13}-n_{23}+1)(n_{13}-n_{33}+3)(n_{13}-n_{33}+2)}} \nonumber \\
    & \times \sqrt{\frac{(n_{22}-n_{33}+1)(n_{13}-n_{12}+1)(n_{12}-n_{11})}{(n_{12}-n_{22}+1)(n_{12}-n_{22})}} \nonumber 
\end{align}
\begin{align}
    e_{14,23,12}(G) &= -\sqrt{\frac{(n_{14}-n_{23}+1)(n_{24}-n_{23})(n_{23}-n_{34}+1)(n_{23}-n_{44}+2)(n_{13}-n_{12})}{(n_{13}-n_{23}+1)(n_{13}-n_{23})(n_{23}-n_{33}+2)(n_{23}-n_{33}+1)}} \nonumber \\
    & \times \sqrt{\frac{(n_{12}-n_{33}+2)(n_{23}-n_{22}+1)(n_{11}-n_{22}+1)}{(n_{12}-n_{22}+2)(n_{12}-n_{22}+1)}} \nonumber 
\end{align}
\begin{align}
    e_{14,23,22}(G) &= -\sqrt{\frac{(n_{14}-n_{23}+1)(n_{24}-n_{23})(n_{23}-n_{34}+1)(n_{23}-n_{44}+2)(n_{13}-n_{22}+1)}{(n_{13}-n_{23}+1)(n_{13}-n_{23})(n_{23}-n_{33}+2)(n_{23}-n_{33}+1)}} \nonumber \\
    & \times \sqrt{\frac{(n_{22}-n_{33}+1)(n_{12}-n_{23})(n_{12}-n_{11})}{(n_{12}-n_{22}+1)(n_{12}-n_{22})}} \nonumber 
\end{align}
\begin{align}
    e_{14,33,12}(G) &= -\sqrt{\frac{(n_{14}-n_{33}+2)(n_{24}-n_{33}+1)(n_{34}-n_{33})(n_{33}-n_{44}+1)(n_{13}-n_{12})}{(n_{13}-n_{33}+2)(n_{13}-n_{33}+1)(n_{23}-n_{33}+1)(n_{23}-n_{33})}} \nonumber \\
    & \times \sqrt{\frac{(n_{12}-n_{23}+1)(n_{22}-n_{33})(n_{11}-n_{22}+1)}{(n_{12}-n_{22}+2)(n_{12}-n_{22}+1)}} \nonumber 
\end{align}
\begin{align}
    e_{14,33,22}(G) &= \sqrt{\frac{(n_{14}-n_{33}+2)(n_{24}-n_{33}+1)(n_{34}-n_{33})(n_{33}-n_{44}+1)(n_{13}-n_{22}+1)}{(n_{13}-n_{33}+2)(n_{13}-n_{33}+1)(n_{23}-n_{33}+1)(n_{23}-n_{33})}} \nonumber \\
    & \times \sqrt{\frac{(n_{23}-n_{22})(n_{12}-n_{33}+1)(n_{12}-n_{11})}{(n_{12}-n_{22}+1)(n_{12}-n_{22})}} \nonumber
\end{align}

\section{Action of the U(4) lowering generators}
\label{appendB}
\begin{align}
    e_{21}(G) &= \sqrt{(n_{12}-n_{11}+1)(n_{11}-n_{22})}, \nonumber 
\end{align}
\begin{align}
    e_{32,12}(G) &= \sqrt{\frac{(n_{13}-n_{12}+1)(n_{12}-n_{23})(n_{12}-n_{33}+1)(n_{12}-n_{11})}{(n_{12}-n_{22}+1)(n_{12}-n_{22})}} \nonumber 
\end{align}
\begin{align}
    e_{32,22}(G) &= \sqrt{\frac{(n_{13}-n_{22}+2)(n_{23}-n_{22}+1)(n_{22}-n_{33})(n_{11}-n_{22}+1)}{(n_{12}-n_{22}+2)(n_{12}-n_{22}+1)}},  \nonumber 
\end{align}
\begin{align}
    e_{31,12}(G) &= \sqrt{\frac{(n_{13}-n_{12}+1)(n_{12}-n_{23})(n_{12}-n_{33}+1)(n_{11}-n_{22})}{(n_{12}-n_{22}+1)(n_{12}-n_{22})}}, \nonumber
\end{align}
\begin{align}
    e_{31,22} (G) &= - \sqrt{\frac{(n_{13}-n_{22}+2)(n_{23}-n_{22}+1)(n_{22}-n_{33})(n_{12}-n_{11}+1)}{(n_{12}-n_{22}+2)(n_{12}-n_{22}+1)}}, \nonumber 
\end{align}
\begin{align}
    e_{43,13}(G) &= \sqrt{\frac{(n_{14}-n_{13}+1)(n_{13}-n_{24})(n_{13}-n_{34}+1)(n_{13}-n_{44}+2)(n_{13}-n_{12})(n_{13}-n_{22}+1)}{(n_{13}-n_{23}+1)(n_{13}-n_{23})(n_{13}-n_{33}+2)(n_{13}-n_{33}+1)}} \nonumber 
\end{align}
\begin{align}
    e_{43,23}(G) &= \sqrt{\frac{(n_{14}-n_{23}+2)(n_{24}-n_{23}+1)(n_{23}-n_{34})(n_{23}-n_{44}+1)(n_{12}-n_{23}+1)(n_{23}-n_{22})}{(n_{13}-n_{23}+2)(n_{13}-n_{23}+1)(n_{23}-n_{33}+1)(n_{23}-n_{33})}} \nonumber 
\end{align}
\begin{align}
    e_{43,33}(G) &= \sqrt{\frac{(n_{14}-n_{33}+3)(n_{24}-n_{33}+2)(n_{34}-n_{33}+1)(n_{33}-n_{44})(n_{12}-n_{33}+2)(n_{22}-n_{33}+1)}{(n_{13}-n_{33}+3)(n_{13}-n_{33}+2)(n_{23}-n_{33}+2)(n_{23}-n_{33}+1)}} \nonumber 
\end{align}
\begin{align}
    e_{42,13,12}(G) &= \sqrt{\frac{(n_{14}-n_{13}+1)(n_{13}-n_{24})(n_{13}-n_{34}+1)(n_{13}-n_{44}+2)(n_{12}-n_{23})}{(n_{13}-n_{23}+1)(n_{13}-n_{23})(n_{13}-n_{33}+2)(n_{13}-n_{33}+1)}} 
    \nonumber \\
    & \times \sqrt{\frac{(n_{12}-n_{33}+1)(n_{13}-n_{22}+1)(n_{12}-n_{11})}{(n_{12}-n_{22}+1)(n_{12}-n_{22})}} \nonumber 
\end{align}
\begin{align}
    e_{42,13,22}(G) &= \sqrt{\frac{(n_{14}-n_{13}+1)(n_{13}-n_{24})(n_{13}-n_{34}+1)(n_{13}-n_{44}+2)(n_{23}-n_{22}+1)}{(n_{13}-n_{23}+1)(n_{13}-n_{23})(n_{13}-n_{33}+2)(n_{13}-n_{33}+1)}} \nonumber \\
    & \times \sqrt{\frac{(n_{22}-n_{33})(n_{13}-n_{12})(n_{11}-n_{22}+1)}{(n_{12}-n_{22}+2)(n_{12}-n_{22}+1)}} \nonumber 
\end{align}
\begin{align}
    e_{42,23,12}(G) &= -\sqrt{\frac{(n_{14}-n_{23}+2)(n_{24}-n_{23}+1)(n_{23}-n_{34})(n_{23}-n_{44}+1)(n_{13}-n_{12}+1)}{(n_{13}-n_{23}+2)(n_{13}-n_{23}+1)(n_{23}-n_{33}+1)(n_{23}-n_{33})}} \nonumber \\
    & \times \sqrt{\frac{(n_{12}-n_{33}+1)(n_{23}-n_{22})(n_{12}-n_{11})}{(n_{12}-n_{22}+1)(n_{12}-n_{22})}} \nonumber 
\end{align}
\begin{align}
    e_{42,23,22}(G) &= \sqrt{\frac{(n_{14}-n_{23}+2)(n_{24}-n_{23}+1)(n_{23}-n_{34})(n_{23}-n_{44}+1)(n_{13}-n_{22}+2)}{(n_{13}-n_{23}+2)(n_{13}-n_{23}+1)(n_{23}-n_{33}+1)(n_{23}-n_{33})}} \nonumber \\
    & \times \sqrt{\frac{(n_{22}-n_{33})(n_{12}-n_{23}+1)(n_{11}-n_{22}+1)}{(n_{12}-n_{22}+2)(n_{12}-n_{22}+1)}} \nonumber 
\end{align}
\begin{align}
    e_{42,33,12}(G) &= -\sqrt{\frac{(n_{14}-n_{33}+3)(n_{24}-n_{33}+2)(n_{34}-n_{33}+1)(n_{33}-n_{44})(n_{13}-n_{12}+1)}{(n_{13}-n_{33}+3)(n_{13}-n_{33}+2)(n_{23}-n_{33}+2)(n_{23}-n_{33}+1)}} \nonumber \\
    & \times \sqrt{\frac{(n_{12}-n_{23})(n_{22}-n_{33}+1)(n_{12}-n_{11})}{(n_{12}-n_{22}+1)(n_{12}-n_{22})}} \nonumber 
\end{align}
\begin{align}
    e_{42,33,22}(G) &= -\sqrt{\frac{(n_{14}-n_{33}+3)(n_{24}-n_{33}+2)(n_{34}-n_{33}+1)(n_{33}-n_{44})(n_{13}-n_{22}+2)}{(n_{13}-n_{33}+3)(n_{13}-n_{33}+2)(n_{23}-n_{33}+2)(n_{23}-n_{33}+1)}} \nonumber \\
    & \times \sqrt{\frac{(n_{23}-n_{22}+1)(n_{12}-n_{33}+2)(n_{11}-n_{22}+1)}{(n_{12}-n_{22}+2)(n_{12}-n_{22}+1)}} \nonumber
\end{align}
\begin{align}
    e_{41,13,12}(G) &= \sqrt{\frac{(n_{14}-n_{13}+1)(n_{13}-n_{24})(n_{13}-n_{34}+1)(n_{13}-n_{44}+2)(n_{12}-n_{23})}{(n_{13}-n_{23}+1)(n_{13}-n_{23})(n_{13}-n_{33}+2)(n_{13}-n_{33}+1)}} 
    \nonumber \\
    & \times \sqrt{\frac{(n_{12}-n_{33}+1)(n_{13}-n_{22}+1)(n_{11}-n_{22})}{(n_{12}-n_{22}+1)(n_{12}-n_{22})}} \nonumber 
\end{align}
\begin{align}
    e_{41,13,22}(G) &= -\sqrt{\frac{(n_{14}-n_{13}+1)(n_{13}-n_{24})(n_{13}-n_{34}+1)(n_{13}-n_{44}+2)(n_{23}-n_{22}+1)}{(n_{13}-n_{23}+1)(n_{13}-n_{23})(n_{13}-n_{33}+2)(n_{13}-n_{33}+1)}} \nonumber \\
    & \times \sqrt{\frac{(n_{22}-n_{33})(n_{13}-n_{12})(n_{12}-n_{11}+1)}{(n_{12}-n_{22}+2)(n_{12}-n_{22}+1)}} \nonumber 
\end{align}
\begin{align}
    e_{41,23,12}(G) &= -\sqrt{\frac{(n_{14}-n_{23}+2)(n_{24}-n_{23}+1)(n_{23}-n_{34})(n_{23}-n_{44}+1)(n_{13}-n_{12}+1)}{(n_{13}-n_{23}+2)(n_{13}-n_{23}+1)(n_{23}-n_{33}+1)(n_{23}-n_{33})}} \nonumber \\
    & \times \sqrt{\frac{(n_{12}-n_{33}+1)(n_{23}-n_{22})(n_{11}-n_{22})}{(n_{12}-n_{22}+1)(n_{12}-n_{22})}} \nonumber 
\end{align}
\begin{align}
    e_{41,23,22}(G) &= -\sqrt{\frac{(n_{14}-n_{23}+2)(n_{24}-n_{23}+1)(n_{23}-n_{34})(n_{23}-n_{44}+1)(n_{13}-n_{22}+2)}{(n_{13}-n_{23}+2)(n_{13}-n_{23}+1)(n_{23}-n_{33}+1)(n_{23}-n_{33})}} \nonumber \\
    & \times \sqrt{\frac{(n_{22}-n_{33})(n_{12}-n_{23}+1)(n_{12}-n_{11}+1)}{(n_{12}-n_{22}+2)(n_{12}-n_{22}+1)}} \nonumber 
\end{align}
\begin{align}
    e_{41,33,12}(G) &= -\sqrt{\frac{(n_{14}-n_{33}+3)(n_{24}-n_{33}+2)(n_{34}-n_{33}+1)(n_{33}-n_{44})(n_{13}-n_{12}+1)}{(n_{13}-n_{33}+3)(n_{13}-n_{33}+2)(n_{23}-n_{33}+2)(n_{23}-n_{33}+1)}} \nonumber \\
    & \times \sqrt{\frac{(n_{12}-n_{23})(n_{22}-n_{33}+1)(n_{11}-n_{22})}{(n_{12}-n_{22}+1)(n_{12}-n_{22})}} \nonumber 
\end{align}
\begin{align}
    e_{41,33,22}(G) &= \sqrt{\frac{(n_{14}-n_{33}+3)(n_{24}-n_{33}+2)(n_{34}-n_{33}+1)(n_{33}-n_{44})(n_{13}-n_{22}+2)}{(n_{13}-n_{33}+3)(n_{13}-n_{33}+2)(n_{23}-n_{33}+2)(n_{23}-n_{33}+1)}} \nonumber \\
    & \times \sqrt{\frac{(n_{23}-n_{22}+1)(n_{12}-n_{33}+2)(n_{12}-n_{11}+1)}{(n_{12}-n_{22}+2)(n_{12}-n_{22}+1)}} \nonumber
\end{align}

\end{document}